\documentclass[12pt,a4paper]{article}

\usepackage{pstricks}
\usepackage{pst-slpe}
\usepackage{pst-blur}

\usepackage{amsfonts}
\usepackage[intlimits]{amsmath}
\usepackage{pdfcolmk}
\usepackage{mathrsfs}
\usepackage{booktabs}
\usepackage{fancyhdr}
\usepackage{epsfig}
\usepackage{verbatim}
\usepackage{color}

\usepackage{lscape}

\newcommand{\E}{{\cal E}}
\renewcommand{\S}{{\cal S}}
\newcommand{\beq}{\begin{equation}}
\newcommand{\eeq}{\end{equation}}
\newcommand\beqa{\begin{eqnarray}}
\newcommand\eeqa{\end{eqnarray}}
\newcommand\bea{\begin{array}}
\newcommand\eea{\end{array}}
\newcommand{\nn}{\nonumber}
\newcommand{\neqa}{\nonumber\end{eqnarray}}
\newcommand{\la}{\label}
\newcommand{\J}{{\cal J}}

\renewcommand{\O}{{\cal O}}

\newcommand{\res}[1]{
\bea{c}
\\[0.05cm]
{\rm res}\\[-0.1cm]
^{#1}
\eea}
\newcommand{\eq}[1]{(\ref{#1})}
\newcommand{\eqs}[2]{(\ref{#1},\ref{#2})}

\newcommand{\h}{\hat}
\renewcommand{\t}{\tilde}

\def\({\left(}
\def\){\right)}
\def\[{\left[}
\def\]{\right]}

\def\<{\langle}
\def\>{\rangle}

\def\d{\partial}
\def\str{{\rm str\,}}

\begin{document}

\begin{flushright}
LPTENS-07/14\\
hep-th/0703191
\end{flushright}
\vspace{1cm}
\begin{center}
{\Large\bf The $AdS_5\times S^5$ superstring quantum spectrum from the algebraic curve}\\
\vspace{1cm}
Nikolay Gromov\footnote{nikgromov@gmail.com}$^{a,b}$,\hspace{0.3cm}
Pedro Vieira\footnote{pedrogvieira@gmail.com}$^{a,c}$\\
\vspace{1cm}
{\it\footnotesize $^a$ Laboratoire de Physique Th\'eorique
de l'Ecole Normale Sup\'erieure\\ et l'Universit\'e Paris-VI,
Paris, 75231, France\\ \vspace{.2cm}
$^b$ St.Petersburg INP, Gatchina, 188 300, St.Petersburg, Russia
\\ \vspace{.2cm}
$^c$
 Departamento de F\'\i sica e Centro de F\'\i sica do Porto
Faculdade de Ci\^encias da Universidade do Porto\\
Rua do Campo Alegre, 687, \,4169-007 Porto, Portugal }

\end{center}
\vspace{1cm}
\begin{abstract}
We propose a method for computing the energy level spacing around classical string solutions in $AdS_5\times S^5$. This method is based on the integrable structure of the string and can be applied to an arbitrary classical configuration.
Our approach treats in equal footing the bosonic and fermionic excitations and provides an unambiguous prescription for the labeling of the fluctuation frequencies.
Finally we revisit the computation of these frequencies for the $SU(2)$ and $SL(2)$ circular strings and compare our results to the existing ones.

\end{abstract}
\pagestyle{empty}
\newpage
\pagestyle{plain}
\tableofcontents
\newpage

\section{Introduction }

In this paper we will study the semi-classical quantization of the $AdS_5\times S^5$ Metsaev-Tseytlin
superstring \cite{Metsaev:1998it}. As a warm-up let us consider a one-dimensional non-relativistic particle in a smooth potential. In terms of the
quasi-momenta
$$
 p(x)\equiv  \frac{\hbar}{i} \frac{\psi'(x)}{\psi(x)}\,,
$$
the Schrodinger equation for the wave function $\psi$ takes the Riccati form
\beq
p^2-i\hbar p'=2m\(E-V\) \,. \la{Schrodinger}
\eeq
What do we know about $p(x)$? It is an analytical function which has, by definition, a pole with residue
\beq
\alpha=\frac{\hbar}{i}  \la{alphaS}
\eeq
at each of the zeros of the wave function. For the $N$-th excited state we will have $N$ poles. On the other hand, for very excited states, the right hand side in (\ref{Schrodinger}) is much larger than $\hbar$ and
\beq
p\simeq p_{{\rm cl} } \equiv \sqrt{2m\(E-V\)} \, \nn
\eeq
describes now a two-sheet Riemann surface. What happened was that, as $N\rightarrow \infty$, the poles in $p(x)$ started to be denser and denser, condensing in a square root cut. Thus, in the semiclassical limit we retrieve the Bohr-Sommerfeld quantization
\beq
\frac{1}{2\pi\hbar}\oint_{\mathcal{C}}  \,p_{{\rm cl}}(z) \,dz\simeq \frac{1}{2\pi\hbar}\oint_{\mathcal{C}}  \,p(z) \,dz=N  \,, \la{bohr}
\eeq
where $\mathcal{C}$ encircles the cut. The first integral is precisely the action variable of the classical motion. To anticipate the forthcoming notation we name such integrals \textit{filling fractions}.

 \vspace{0.5cm}
\begin{figure}[h]
    \centering
        \resizebox{160mm}{!}{\includegraphics{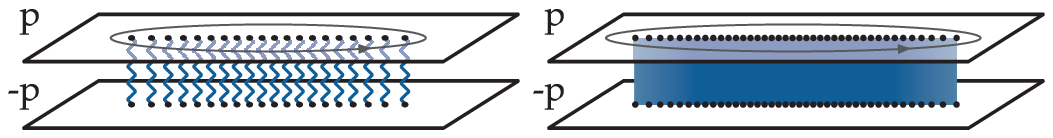}}
    \caption{\small\textsf{\textit{Analytical structure of a quasi-momenta $p(x)$ of a
    one dimensional system. Left: for low
    lying states $p(x)$ is a collection of poles. Right: for high energy states
    the poles condense into a square root branch cut.}}\label{fig:simple}
}
\end{figure}

When we consider more degrees of freedom, in particular when we move to higher dimensions, let us say two, the situation is
not just a little worse. Indeed, we have no proper \textit{generic} recipe, except from lattice calculations, to extract the quantum
spectrum, or a part of it, of an interacting quantum field theory. However, if we are lucky, it might happen that the theory is
integrable. If it is the case, we can identify the action variables, apply the Bohr-Sommerfield condition and find the quasi-classical spectrum of the theory.

For a wide class of two dimensional sigma models this happens to be the case and the procedure is known explicitly. The central object is a collection of quasi-momenta, $p_i(x)$, whose derivative defines a many-sheet Riemann surface. These sheets can be connected by several cuts, to each of which we can associate a \textit{filling fraction}
 by integrating the quasi-momenta around the cut as in (\ref{bohr}). These are the action variables of the theory.  Grosso modo, these filling fractions measure the size of the cut. Finally, when going through these cuts the quasi-momenta can jump by $ 2\pi n$ with $n$ being an integer labeling the cut.

The superstring on $AdS_5\times S^5$ background falls into this class of theories -- the model is known to be classically integrable \cite{Mandal:2002fs,Bena:2003wd}, the algebraic curve was built \cite{BKSZ}, and thus one can expect to quasi-classically quantize the string. In the string language, when we choose which Riemann sheets we connect by a cut we are choosing which string polarization, i.e. which degree of freedom, to excite. The number $n$ and the filling fraction associated to the cut are in strict analogy with the mode number and amplitude of a Fourier mode in a free theory such as the string in flat space \cite{BKSZ}.

It is now widely believed that the integral equations describing the  quasi-momenta $p_i(x)$ are the continuous limit of a system of discrete equations carrying the name of Bethe ansatz equations \cite{Arutyunov:2004vx,Beisert:2005fw}. As in the example we considered, the quasi-momenta are given by a sum of poles which condense into cuts in the classical limit \cite{KMMZ}.
\begin{figure}[t]
\label{fig:ex}
    \centering
        \resizebox{120mm}{!}{\includegraphics{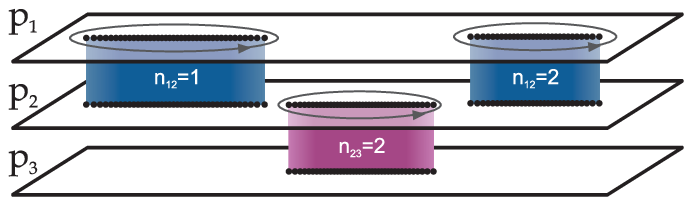}}
    \caption{\small\textsf{\textit{
    A possible analytical stricture of the quasi-momenta
    of an integrable sigma model. Many types of cuts are now possible.
    Cuts can join different sheets and each cut is marked by its ``mode
    number" $n_{ij}$. In flat space limit they become numbers of fourier modes. The number
    of microscopical poles constituting the given cut is called a ``filling fraction" and can
    be calculated as a contour integral \eq{bohr}.}}}
\end{figure}

Going back to our simple example, we can see that the existence of such
discrete equations is indeed highly natural. For that purpose let us
consider a simple harmonic oscillator, $V=\frac{m\,\omega^2 x^2}{2}$.
From (\ref{Schrodinger}) it follows that $p(x)=im \omega
x+\mathcal{O}(1/x)$. Since the quasi-momentum is a meromorphic function
with $N$ poles on the real axis, it must be given by
\beq
\nn p(x)=im \omega x+\frac{\hbar}{i}\sum_{i=1}^{N}\frac{1}{x-x_i} \,.
\eeq
Then, from the large $x$ behavior in (\ref{Schrodinger}) we read immediately
\beq
E=\hbar\omega\(N+\frac{1}{2}\) \nn
\eeq
while from the cancelation of each of the $x_i$ poles in the same equation we get\footnote{Its solution is given by the zeros of the Hermite polynomials,
$
H_N\(\sqrt{\frac{2m\omega}{\hbar}} x_i\)=0\,.
$}
\beq
x_i=\frac{\hbar}{2\omega m}\sum_{j\neq i}^N \frac{1}{x_i-x_j} \la{Hermite}
\eeq
which strongly resembles the equations one finds in the Bethe ansatz context.

At this point let us leave this instructive example behind and resume our main discussion.
When we expand the superstring action around some classical solution, characterized by some conserved charges, we obtain, for the oscillations, a quadratic lagrangian whose quantization yields, for the semiclassical spectrum,
\beq
E=E_{cl} +\sum_{A,n} N_{A,n} \,\E_{A,n}\, , \la{qtE}
\eeq
where we have dropped the zero energy excitation and denoted the number of quanta with energy $\E_{A,n}$ by $N_{A,n}$. The subscript $A$ labels the several possible string polarizations we can excite while the mode number $n$ is the Fourier mode of the quantum fluctuation. In this article we shall address the question of finding this quasi-classical spectrum for the $AdS_5 \times S^5$ superstring using the algebraic curve mentioned above.

Let us explain the idea behind the computation. There are basically two main steps involved. First we construct the curve associated with the classical trajectory around which we want to consider the quantum fluctuations. This will be given by some Riemann surface with some cuts uniting some of the eight sheets. The second step consist of considering the small excitations around this classical solution in the spirit of \cite{Beisert:2005bv}. In terms of the algebraic curve this consists of adding some
microscopic cuts to this macroscopic background. By microscopic cuts we mean some finite number of poles whose residue we know \cite{Arutyunov:2004vx,Beisert:2005fw,GKSV}, just like in the simple example (\ref{alphaS}). Then, by construction, the energy of the perturbed configuration is quantized as in (\ref{qtE}). We must stress that the knowledge of the residue, of utmost importance, is indeed the only extra input we needed to compute the quasi-classical spectrum.

As an application of this method we compute the fluctuation frequencies around the circular $su(2)$ and $sl(2)$ string. These solutions belong to a family of circular solutions whose quasi-momenta we computed explicitly in Appendix A. The frequencies we compute in this way were obtained in \cite{Frolov:2003qc,Frolov:2003tu} and \cite{Arutyunov:2003za,Park:2005ji} by direct analysis of the fluctuations in the Metsaev-Tseytlin GS superstring action.
In the end of section 5 we will propose a precise prescription for the labeling of the fluctuation frequencies and discuss briefly the computation of the $1$--loop shift.

\section{Coset Model and Algebraic Curve}\label{Coset}

In this section we shall review the construction of the algebraic curve of Beisert, Kazakov, Sakai and Zarembo \cite{BKSZ} describing the classical motion of the superstring in the target superspace
$$\frac{PSU(2,2|4)}{SP(2,2)\times SP(4)}$$
whose bosonic part is $ AdS_5\times S^5 $. The Metsaev-Tseytlin superstring action is of GS type, formulated as a supercoset \cite{Metsaev:1998it}.

\subsection{The supercoset}
This section follows closely \cite{Alday:2005jm}.
The matrix superalgebra $su(2,2|4)$ is spanned by the $8\times 8$ supertraceless supermatrices
\beq
M=\( \begin{array}{c|c}
A & B \\ \midrule
C & D
\end{array}\) \nn
\eeq
where $A$ and $B$ belong to $u(2,2)$ and $u(4)$ respectively while the fermionic components are related by
$$ C=B^\dagger \( \bea{cc}
{\mathbb I}_{2\times2}&0\\
 0&-{\mathbb I}_{2\times2}
\eea\)\,. $$
The $psu(2,2|4)$ superalgebra is the quotient of this algebra by the matrices proportional to the identity.
Since the $su(2,2|4)$ algebra enjoys the automorphism
\beq
\Omega \circ M=\(\bea{cc}EA^TE&-EC^TE\\ E B^T E&ED^T E\eea\) \,\, , \,\, E=\( \bea{cccc}
 0&-1&0&0\\
 1&0&0&0\\
 0&0& 0&-1\\
 0&0&1&0
\eea\)\, , \nn
\eeq
such that $\Omega^4=1$, the algebra is endowed with a $\mathbb Z_4$ grading.
This means that any algebra element can be decomposed into $\sum_{i=0}^3M^{(i)}$,
where $\Omega\circ M^{(n)}=i^n M^{(n)}$. Explicitly
\beqa
\bea{ccc}
M^{(0,2)} &=&\frac{1}{2}\(\bea{cc}A\pm E A^T E&0\\0&D\pm E D^T E\eea\)\\
\midrule
M^{(1,3)}&=&\frac{1}{2}\(\bea{cc}0&B\pm iE C^T E\\C\mp i E B^T E&0\eea\) \\
\eea  \,.\label{Mns}
\eeqa
We see that the $M^{(0)}$ elements span, by definition, the denominator algebra
$
sp(2,2)\times sp(4)
$
of the coset.
Then, the remaining bosonic elements, $M^{(2)}$, orthogonal to the former, generate the (orthogonal) complement of $sp(2,2)\times sp(4)$ in $su(2,2)\times su(4)$.

Finally, the Metsaev-Tseytlin action for the GS superstring in $AdS_5\times S^5$ is then given in terms of the algebra current
\beq
J=-g^{-1}dg\,, \label{J}
\eeq
where $g(\sigma,\tau)$ is a group element of $PSU(2,2|4)$, by
\beq
S=\frac{\sqrt{\lambda}}{4\pi}\int \str \(J^{(2)}\wedge  *J^{(2)}-J^{(1)}\wedge J^{(3)}\) +\Lambda \wedge \str J^{(2)} \, ,
\label{ActionJ}
\eeq
where the last term ensures that $J^{(2)}$ is supertraceless. 
Besides the obvious global $PSU(2,2|4)$ left multiplication symmetry the action (\ref{ActionJ})
possesses a local gauge symmetry, $g\rightarrow gH$ with $H \in SP(2,2)\times SP(4)$, under which
\beqa
J^{(i)}&\rightarrow& H^{-1}J^{(i)}H \,\,\quad , \,\,\quad i=1,2,3\, \nn
\eeqa
while $J^{(0)}$ transforms as a connection.
The equations of motion following from (\ref{ActionJ}) are equivalent to the conservation of the Noether current associated with the global left multiplication symmetry
\beq
d*k=0  \label{eomk}
\eeq
where $k=gKg^{-1}$ and $K=J^{(2)}+\frac{1}{2}*J^{(1)}-\frac{1}{2}*J^{(3)}-\frac{1}{2}*\Lambda $.

For a purely bosonic representative $g$ we can write
\beqa
g=\(\bea{c|c}
\mathcal{Q} &\;\;\;0\;\;\;\\ \midrule
\;\;\;0\;\;\;& \mathcal{R}
\eea\) \nn \,.
\eeqa
where $\mathcal{R} \in SU(4)$ and $\mathcal{Q}\in SU(2,2)$. Then we see that
$\mathcal{R} E \mathcal{R}^T  $
is a good parametrization of
$$SU(4)/SP(4) =S^5 $$
because, by definition, it is invariant under $\mathcal{R}\rightarrow \mathcal{R} H$ with $H\in SP(4)$. In the same way $\mathcal{Q} E \mathcal{Q}^T $ describes the $AdS$ space. Then we can define the embedding coordinates $u$ and $v$ by the simple relations
\beq
u^j\Sigma^{S}_j =\mathcal{R} E \mathcal{R}^T  \,\, , \,\, v^j\Sigma^{A}_j=\mathcal{Q} E \mathcal{Q}^T \la{map}
\eeq
where $\Sigma^{S},\Sigma^A$ are the gamma matrices of $SO(6)$ and $SO(4,2)$. By construction these coordinates will automatically satisfy
\beqa
1&=&  u_6^2+u_5^2+u_4^2+u_3^3+u_2^2+u_1^2\,, \nn \\
1&=&  v_6^2+v_5^2- v_4^2- v_3^2- v_2^2-   v_1^2\,. \la{const}
\eeqa
Then the bosonic part of the action can be expressed in the usual non--linear $\sigma$ model form
\beqa
S_b=\frac{\sqrt{\lambda}}{4\pi}\int_0^{2\pi} d\sigma \int d\tau\sqrt{h}\, \(h^{\mu\nu}\, \partial_\mu u\cdot\partial_\nu u +\,\lambda_{u} \(u\cdot u-1\) - (u\rightarrow v)\) \, . \nn
\eeqa

\subsection{The algebraic curve}\label{curve}
As follows from the equations of motion and the flatness condition,
\beq
dJ-J\wedge J=0\,,   \nn
\eeq
the connection
\beq
A(x)=J^{(0)}+\frac{x^2+1}{x^2-1}\,J^{(2)}
-\frac{2x}{x^2-1}\(*J^{(2)}-\Lambda\)+\sqrt{\frac{x+1}{x-1}} \,J^{(1)}+\sqrt{\frac{x-1}{x+1}} \,J^{(3)} \label{Ax}
\eeq
is flat for any complex number $x$ \cite{Bena:2003wd}. This is the crucial observation which indicates the model to be (at least classically) integrable. Indeed, we can define the monodromy matrix
\beq
\Omega(x)={\rm Pexp\,}\oint_\gamma A(x) \label{monodromy}
\eeq
where $\gamma$ is any path starting and ending at some point $(\sigma,\tau)$ and wrapping the worldsheet cylinder once. Since the flatness of the connection ensures path independence we can choose $\gamma$ to be the constant $\tau$ path. Moreover, placing this loop at some other value of $\tau$ just amounts to a similarity transformation of the monodromy matrix. Thus we conclude that the eigenvalues of $\Omega(x)$ are time independent. Since they depend on a generic complex number $x$, we have obtained in this way an infinite number of conserved charges thus assuring integrability. Finally we notice that under periodic $SP(2,2)\times SP(4)$ gauge transformations the monodromy matrix transforms by a simple similarity transformation so that the eigenvalues are also gauge invariant.

In the rest of this section we shall review the results of \cite{BKSZ} and analyze the analytical properties the quasi-momenta $\h p$ and $\t p$ defined from the eigenvalues
\beq
\{e^{i\h p_1},e^{i\h p_2},e^{i\h p_3},e^{i\h p_4}|e^{i\t p_1},e^{i\t p_2},e^{i\t p_3},e^{i\t p_4} \} \nn
\eeq
of the monodromy matrix $\Omega(x)$.
The eigenvalues are the roots of the characteristic polynomial equation and thus they define an 8--sheet  Riemann surface. These sheets are connected by several cuts\footnote{This connection appears as square root cuts, if the cut relates $\h p$'s (of $\t p$'s) among themselves, or by poles, if the pole is shared between some $\h p_i$ and $\t p_j$ \cite{BKSZ}.} -- see fig. 2 -- whose branchpoints are the loci  where the eigenvalues of the monodromy matrix become equal. The quasi-momenta can jump by a multiple of $2\pi $ at points connected by a cut\footnote{Note that the derivative of the quasi-momenta is a single valued function on the Riemann surface while $p(x)$ is not.}. For example,
for a cut going from the first to the second sheet , we will have
\beq
\h p_1^+-\h p_2^-=2\pi n \nn \,\, , \,\, x\in \mathcal{C}_{n}^{\h1 \h2}
\eeq
where $\t p^\pm$ stands for the value of the quasi-momenta immediately above/below the cut. This integer $n$, together with the filling fraction we shall introduce in next section, label each of the cuts.
Generically, we can summarize all equations as
\beq
p_i^+- p_j^-=2\pi n_{ij} \,\, , \,\, x\in \mathcal{C}_{n}^{ij} \label{n32}
\eeq
where the indices $i$ and $j$ take values
\beq
\la{range} i=\t 1,\t 2,\h 1,\h 2 \,\, , \,\, j=\t 3,\t 4,\h 3,\h 4
\eeq
and we denote
\beq
p_{\t 1,\t 2,\t 3,\t 4}\equiv \t p_{1,2,3,4} \,\, , \,\, p_{\h 1,\h 2,\h 3,\h 4}\equiv \h p_{1,2,3,4} \,. \la{notation}
\eeq
In section \ref{FRE} we will explain why we should consider only these $16$ possible excitations. Moreover to each cut we also associate the filling fraction
\beq
S_{ij}=\pm\,\frac{\sqrt{\lambda}}{8\pi^2i}\oint_{\mathcal{C}_{ij}} \(1-\frac{1}{x^2}\) p_i(x) dx \label{filling}.
\eeq
obtained by integrating the quasi-momenta around the square root cut. As before, the indices run over (\ref{range}) and we should chose the plus sign for $i=\h 1,\h 2$ and the minus sign for the remaining excitations with $i=\t 1,\t 2$.
Let us explain why we chose to integrate the quasi-momenta $p(x)$ around the cut with the seemingly mysterious $1-1/x^2$ weight.
In terms of the Zhukovsky variable $z=x+1/x$ equation (\ref{filling}) this expression takes the standard form \eq{bohr}.
It was pointed out in
\cite{Beisert:2004ag,BKSZ} and shown in
\cite{Dorey:2006zj} that these filling fractions are indeed the action variables of the theory.  Indeed, from the AdS/CFT
correspondence these filling fractions are expected to be integers
since they correspond to an integer number of Bethe roots on the SYM side
\cite{KMMZ,Beisert:2005di}. Indeed, the likely existence of the Bethe ansatz description \cite{Arutyunov:2004vx,Beisert:2005fw}
of the $AdS_5\times S^5$ superstring also implies this pole structure of the exact quasi-momentum
in a semi-classical limit. Moreover, in \cite{GKSV}, where the $S^5$ subsector was studied it was clear that the quasi-momenta $p(z)$ coming from the quantum Bethe ansatz equations appears in the usual form, with unit residue for each pole, for the Zhukovsky variable $z$.
Thus (\ref{filling}) is the good starting point for the string quasi-classical quantization.

From  (\ref{Mns}),(\ref{Ax}) it follows that
\beq
C^{-1}\,\Omega(x)\,C=\Omega^{-ST}(1/x),\;\;\;\;\;C=
\(
\bea{c|c}
 E&0\\ \midrule
 0&-i E
\eea\) \nn
\eeq
which translates into the inversion symmetry
\beqa\nn
\t p_{1,2}(x)&=&-2\pi m-\t p_{2,1}(1/x)\\ \la{x1/x}
\t p_{3,4}(x)&=&+2\pi m-\t p_{4,3}(1/x)\\
\h p_{1,2,3,4}(x)&=&-\h p_{2,1,4,3}(1/x)\nn
\eeqa
for the quasi-momenta\footnote{Note that for $\hat p$ there is no $2\pi m$ imposed by requiring absence of time windings \cite{Kazakov:2004nh,BKSZ}.}.

The singularities of the connection at $x=\pm 1$ result in simple poles for the quasi-momenta. These singularities come from the current $J^{(2)}$ in (\ref{Ax}).  This current is supertraceless because it belongs to $psu(2,2|4)$ and so is its square due to the Virasoro constraints following from the variation of the action with respect to the worldsheet metric. Together with the inversion symmetry this forces the various residues to organize as follows
 \beq
\{\h p_1,\h p_2,\h p_3,\h p_4|\t p_1,\t p_2,\t p_3,\t p_4\}\simeq
\frac{\{  \alpha_\pm,  \alpha_\pm,  \beta_\pm,  \beta_\pm|  \alpha_\pm,  \alpha_\pm,  \beta_\pm,  \beta_\pm\}}{x\pm 1} \label{pm1} \,.
\eeq
Thus we see that the residues at these points are synchronized and must be the same for the $S^5$ and the $AdS_5$ quasi-momenta $\h p_i$ and $\t p_i$. This is the crucial role of the Virasoro constraints which will be of utmost importance in the remaining of the article.

Finally, for large $x$, one has
\beq
A_\sigma\simeq -g^{-1}\(\d_\sigma+\frac{2}{x}\,k_\tau\)g \nn
\eeq
where $k$, defined bellow (\ref{eomk}), is the Noether current associated with the left global symmetry. Thus, from the behavior at infinity we can read the conserved global charges\footnote{These are the bosonic charges, the ones which are present for a classical solution. Latter we shall consider all kind of fluctuations, including the fermionic ones. Then we shall slightly generalize this expression to (\ref{largex}).}
\beq
\(\bea{c}
\h p_1\\
\h p_2\\
\h p_3\\
\h p_4\\  \midrule
\t p_1\\
\t p_2\\
\t p_3\\
\t p_4\\
\eea\) \simeq  \frac{2\pi }{x\sqrt\lambda}
\(\bea{l}
+E-S_1+S_2 \\
+E+S_1 -S_2 \\
-E-S_1  -S_2 \\
-E+S_1 +S_2 \\  \midrule
+J_1+J_2-J_3  \\
+J_1-J_2+J_3 \\
-J_1+J_2 +J_3 \\
-J_1-J_2-J_3
\eea\)\,. \label{inf}
\eeq

The finite gap method allow us to build, at least implicitly, classical solutions
of the nonlinear equations of motion from the analytical properties of the
quasi-momenta\footnote{For the inverse problem of recovering the solutions from the algebraic curve see the monographs
\cite{2003icis.book.....B,belokos} for the general formalism and \cite{Dorey:2006zj} where this is carried over in the context of string theory for the classical bosonic string in $S^3\times R\subset AdS_5\times S^5$ described by the KMMZ algebraic curve \cite{KMMZ}.}.

As we shall see in the next section, the algebraic curve can also be turned into a powerful tool to study the quantum spectrum, i.e. the energy level spacing, for energies close to that of a given classical string solution.

\section{Circular string solutions}\la{circular}
In this section we will write down an important class of rigid circular strings studied in \cite{Arutyunov:2003za}. As we explain below that are particularly simple from the algebraic curve point of view and will therefore provide us an excellent playground to check our method for some simple choice of parameters.
In terms of the $AdS_5$ and $S^5$ embedding coordinates,
we can represent this general class of strings solutions with global charges $E=\sqrt{\lambda}\,\mathcal{E}$, $J_1=\sqrt{\lambda}\,\J_1$, $\dots$, as \cite{Arutyunov:2003za}
\beqa
u_2+iu_1=\sqrt{\frac{\J_3}{w_3}} \,e^{i\(w_3\tau+m_3\sigma\)} &,& v_2+iv_1= \nn
\sqrt{\frac{\S_2}{{\rm w}_2}} \,e^{i\({\rm w}_2\tau+k_2\sigma\)}\,, \\
u_4+iu_3=\sqrt{\frac{\J_2}{w_2}} \,e^{i\(w_2\tau+m_2\sigma\)} &,& v_4+iv_3=
\sqrt{\frac{\S_1}{{\rm w}_1}} \,e^{i\({\rm w}_1\tau+k_1\sigma\)}\,,\la{43}\\
u_6+iu_5=\sqrt{\frac{\J_1}{w_1}} \,e^{i\(w_1\tau+m_1\sigma\)} &,& v_6+iv_5=
\sqrt{\frac{\E}{\kappa}} \,e^{i\kappa \tau}  \,, \nn
\eeqa
where the equations of motion, Virasoro constraints and (\ref{const}) impose
\beqa
&&1=\sum\limits_{i=1}^3\frac{\J_i}{w_i}  \,\, , \,\, 1=\frac{\E}{\kappa}-\sum\limits_{j=1}^2\frac{\S_j}{{\rm w}_j} \,\, , \,\, 0=\sum\limits_{j=1}^2 k_j\S_j+\sum\limits_{i=1}^3 m_i\J_i \,,\nn\\
&&{\rm w}_j^2=\kappa^2+k_j^2  \,\, , \,\, \kappa^2=\sum\limits_{j=1}^2\S_j\,\frac{2k_j^2}{{\rm w}_j} +\sum\limits_{i=1}^3\J_i\frac{w_i^2+m_i^2}{w_i} \,,\la{constraints2}\\
&& w_i^2=\nu^2+m_i^2\,\, , \,\, \nu^2 \equiv  \sum\limits_{i=1}^3\J_i\frac{w_i^2-m_i^2}{w_i} \,.\nn
\eeqa
As explained in Appendix A, for this family of solutions the representative $g$ can be written as
$$
g=e^{ \varphi_\sigma\sigma+ \varphi_\tau\tau}\cdot g_0
$$
where $\varphi_{\sigma,\tau}$ are linear combinations of Cartan generators and $g_0$ is a constant matrix. Then we see that the current
$$
J=-g^{-1}dg\,,
$$
and therefore also the flat connection $A(x)$ in (\ref{Ax}), are constant matrices! Then the computation of the path order exponential (\ref{monodromy}) is trivial and the quasi-momenta $p(x)$ are simply obtained from the eigenvalues of $\frac{2\pi}{i}A(x)$. For a detailed account see Appendix A.

\section{Frequencies from the algebraic curve}\la{FRE}
\begin{figure}[t]
\centerline
{
 \resizebox{0.6\textwidth}{!}{\includegraphics{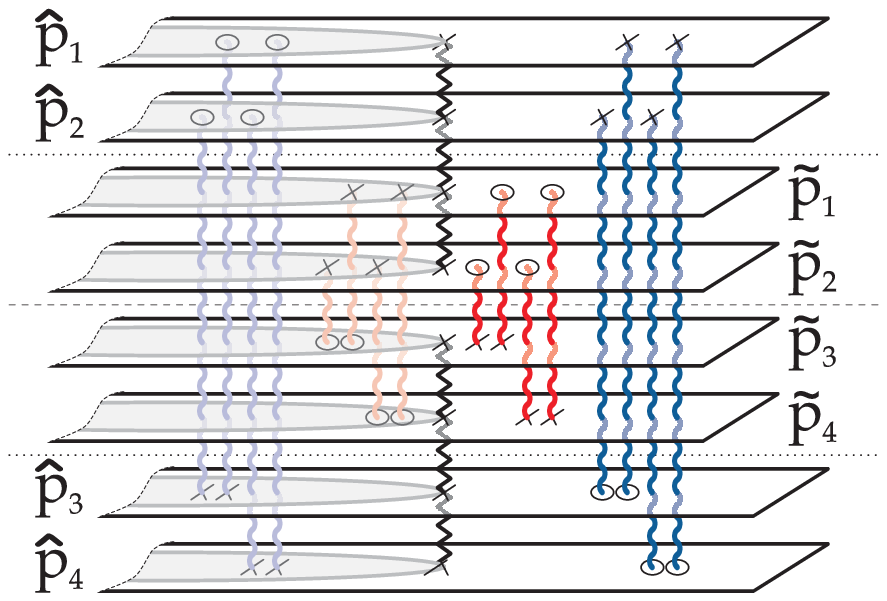}}
}
\caption{\small\textsf{\textit{\label{fig:bp} Some configuration of poles on
the algebraic curve corresponding to the $S^5$ excitations (red) and
$AdS_5$ excitations (blue). Black line denotes poles at $\pm 1$,
connecting 4 sheets with equal residues.  The crosses correspond to the
residue $+\alpha(x)$, while circles to residue $-\alpha(x)$. Physical
domain of the surface lies outside the unit circle.}}}
\end{figure}
\begin{figure}[t]
\centerline
{
 \resizebox{0.6\textwidth}{!}{\includegraphics{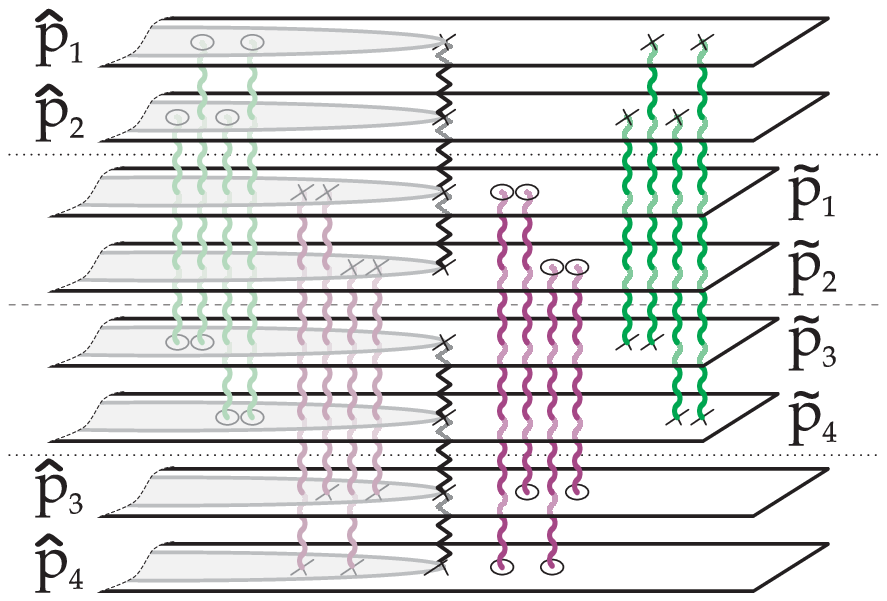}}
} \caption{\small\textsf{\textit{\label{fig:fp} Some configuration of
poles on the algebraic curve corresponding to the 8 fermionic
excitations. Black line denotes poles at $\pm 1$, connecting 4 sheets
with equal residues. The crosses correspond to the residue $\alpha(x)$,
while circles to residue $-\alpha(x)$. Physical domain of the surface
lies outside the unit circle.}}}
\end{figure}
In this section we will consider the
quasi-classical quantization of the $AdS_5\times S^5$ superstring in the language of the algebraic curve. As an example we will
find the low lying energy spectrum for the excitations around some simple classical string solutions.

As we have already mentioned in the introduction and in section \ref{curve} the \textit{exact} quasi-momenta is made out of a large collection of poles. From (\ref{filling}) we infer the residue of each pole,
\beq
p\simeq  \sum_{a=k}^{S_n} \frac{\alpha(x_k)}{x-x_{k}}+\dots \,, \nn
\eeq
with
\beq
\alpha(x)=\frac{4\pi}{\sqrt{\lambda}}\frac{x^2}{x^2-1} \label{alpha} \,.
\eeq
These poles may then condense into square root cuts forming a classical Riemann surface like in fig. 2.
The filling fraction and mode number of the cuts are in strict analogy with the amplitude and mode number of a fourier mode in the usual flat space string. Then, to consider the quantum fluctuations around this classical solution, amounts to adding small cuts, i.e. poles, to this curve.
The key ingredient allowing us to do so is the knowledge of the residue (\ref{alpha}) just like in the example (\ref{alphaS}) in the introduction.
The several possible choices of sheets to be connected by these poles correspond to the several possible polarizations of the
superstring, i.e. to the different quantum numbers. The $16$ physical excitations are the
$4+4$ modes in $AdS_5$ and $S^5$   (fig \ref{fig:bp}) plus the $8$ fermionic fluctuations (fig \ref{fig:fp}).

Let us give a bit more of flavor to the discussion above.
As we mentioned in the introduction, the equations describing the eight sheet quasi-momenta can be discretized \cite{Arutyunov:2004vx} yielding a set of Bethe ansatz  equations for the roots $x_i$ making up the cuts.
The resulting equations resemble (\ref{Hermite}) with an extra $2\pi n_i$ in the left hand side
$$
\sum_{j\neq i}\frac{1}{x_i-x_j}=2\pi n_i+V(x_i) \,.
$$
This means that we can think of $x_i$ as being the position of a particle interacting with many other particles via a two-dimensional Coulomb interaction, placed in an external potential\footnote{in (\ref{Hermite}) the potential is a quadratic one, for the actual Bethe equations it is something else} and feeling an external force $2\pi n_i$. What we are doing is, then, first considering a large number of particles which will condense in some disjoint supports, the cuts, with each cut being made out of particles with the same mode number $n_i$. Then we add an extra particle with some other mode number $n$. At the leading order, two things happen. The particle will seek its equilibrium point in this background and will backreact, shifting this background slightly by its presence \cite{Beisert:2005bv}. The (AdS global time) energy $E$ of the new configuration is then shifted. When adding $N$ particles we get precisely the quantum steps in the spectrum, i.e. (\ref{qtE}).

Technically the computations can be divided into two main steps. In what follows we will use the notation (\ref{notation}) intensively. We must solve (\ref{n32}) for all cuts of the Riemann surface where we now have $p(x)\rightarrow p(x)+\delta p(x)$ where $p(x)$ is the quasi-momenta associated with the classical solution.
\begin{itemize}
\item{
When applied to the microscopic cut, i.e. pole, equation (\ref{n32}) gives us, to leading order, the position $x_n^{ i j}$ of the pole,
\beq
p_{i}(x_n^{ i j})-p_{j}(x_n^{ i j})=2\pi n,\;\;\;\;\;|x_n^{i j}|>1\ , \label{eqsposition}
\eeq
where $i<j$ are taking values $\h 1,\h 2,\h 3,\h 4,\t 1,\t 2,\t 3,\t 4$ and indicate which two sheets share the pole. We refer to domain $|x|>1$ as \textit{physical domain}. The interior of the unit circle is just the mirror image of the physical domain, as we saw in the previous section (\ref{x1/x}).}
\item{
Then, to find $\delta p$, and in particular the energy shift $\delta E$, we must solve the same equations but now in the macroscopic cuts
\beq
\delta p_{i}^+-\delta p_{j}^-=0 \,\, , \,\, x\in \mathcal{C}_n^{ij} \,. \la{problem}
\eeq
This linear problem is to be supplemented with the known analytical properties of $\delta p(x)$ namely the asymptotic behavior presented below and the simple pole singularities with residues (\ref{alpha}). In this way we are computing the backreaction described above.
}
\end{itemize}
Before proceeding  it is useful to introduce some simple notation. We shall consider $N_n^{i j}$ excitations with mode number $n$ between sheet $p_i$ and $p_j$ such that
\beq
N_{i j}\equiv \sum_n N_n^{ i j} \, \nn
\eeq
is the total number of poles connecting these two sheets. Moreover, each excitation has their own quantum numbers according to the global symmetry.
The $S^5$, $AdS_5$ and fermionic excitations can then be identified as the several possible choices of sheets to be connected, see figs 3 and 4,
\beqa
S^5 &,& (i,j)=(\t 1,\t 3),(\t 1,\t 4),(\t 2,\t 3),(\t 2,\t 4) \nn \\
AdS_5 &,& (i,j)=(\h 1,\h 3),(\h 1,\h 4),(\h 2,\h 3),(\h 2,\h 4) \nn \\
\text{Fermions} &,& (i,j)=(\t 1,\h 3),(\t 1,\h 4),(\t 2,\h 3),(\t 2,\h 4), \la{list}\\
&&\,\,\,\,\,\,\,\,\,\,\,\,\,\,\,\,\,\,\ (\h 1,\t 3),(\h 1,\t 4),(\h 2,\t 3),(\h 2,\t 4) \nn
\eeqa
The $16$ physical degrees of freedom of the superstring are precisely these $16$ elementary excitations,
also called  \textit{momentum carrying excitations} \cite{BKSZ,Beisert:2005di}.

When adding extra poles to the classical solutions its energy will be shifted by
\beq
\delta E =\delta \Delta + \sum_{{\rm AdS}^5}N_{ij}+\frac{1}{2}\sum_{{\rm Ferm}}N_{ij}\,, \la{EDelta}
\eeq
where we isolated the anomalous part $\delta \Delta$ of the energy shift from the trivial bare part.
Then, it is convenient to recast (\ref{inf}), for the excitations, as
\beqa
\la{largex}
\delta \(\bea{c}
\h p_1\\
\h p_2\\
\h p_3\\
\h p_4\\  \midrule
\t p_1\\
\t p_2\\
\t p_3\\
\t p_4\\
\eea\) \!\!\! &\simeq& \!\!\!\! \frac{4\pi }{x\sqrt\lambda}\!
\(\bea{rrl}
+\delta \Delta/2&+N_{\h 1\h 4}+N_{\h 1 \h 3}&+N_{\h 1\t 3}+N_{\h1\t 4}\\
+\delta \Delta/2&+N_{\h 2\h 3}+N_{\h 2\h 4}&+N_{\h2\t 4}+N_{\h2\t 3} \\
-\delta \Delta/2&-N_{\h 2\h 3}-N_{\h 1 \h 3}&-N_{\t 1\h3}-N_{\t 2\h3} \\
-\delta \Delta/2&-N_{\h 1\h 4}-N_{\h 2\h 4}&-N_{\t 2\h4}-N_{\t 1\h4} \\ \midrule
&- N_{\t1 \t4}- N_{\t 1\t 3} &-N_{\t 1\h3}-N_{\t 1\h4} \\
&- N_{\t 2 \t3}- N_{ \t 2 \t4}                         &-N_{\t 2\h4}-N_{\t 2\h3}\\
&+ N_{ \t2 \t3}+ N_{ \t1 \t3}&+N_{\h1\t 3}+N_{\h2\t 3}\\
&+ N_{ \t1 \t4}+ N_{ \t 2 \t4}             &+N_{\h2\t 4}+N_{\h1\t 4}
\eea\)
\eeqa
These filling fractions $N^{n}_{ij}$ are not independent. Any algebraic curve must obey the Riemann bilinear identity (see eqs. 3.38 and 3.44 in \cite{BKSZ}). Since this was already the case for the classical solution around which we are expanding, the new filling fractions are constrained by
\beq
\la{hir}\sum_n n \sum_{\rm All \,ij}N_n^{i j}=0 \,,  \eeq
which is nothing but the string level matching condition in the algebraic curve language.

It is also important to note that sign of the residues can be summarized by the following formula\\[-0.9cm]
\beqa
\la{xnres}\res{x=x^{ij}_n} \h p_{k} =\(\delta_{i\h k}-\delta_{j\h k}\)\alpha(x^{ij}_n)N^{ij}_n,\;\;\;\;\;\res{x=x^{ij}_n} \t p_{k} =\(\delta_{j\t k}-\delta_{i\t k}\)\alpha(x^{ij}_n)N^{ij}_n \,,
\eeqa
with $k=1,2,3,4$ and $i<j$ taking values $\h 1,\h 2,\h 3,\h 4,\t 1,\t 2,\t 3,\t 4$, as summarized in figs.\ref{fig:bp} and \ref{fig:fp}.

In the following sections we shall analyze the quantum fluctuations around some simple classical solutions belonging to the family of rigid circular strings (\ref{43}). We will do it in three main steps. First we compute the quasi-momenta\footnote{Due to the simplicity of these solutions we could have computed the quasi-momenta by an alternative method, namely using just the analytical properties presented in section \ref{curve}. This was done for the $su(2)$ and $sl(2)$ circular solutions in \cite{KMMZ} and \cite{Kazakov:2004nh}.} associated to each classical solution as explained in section \ref{circular} and in greater detail in Appendix A.
Then we shall consider the fluctuations around the classical solution which appear as new poles in the quasi-momenta. As explained above, we start by finding the position of these new roots using (\ref{eqsposition}) and then we shall compute the perturbation $\delta p$ of the quasi-momenta by using, again, the analytical properties described in section \ref{curve} plus the knowledge of the poles' positions found in the second step.

We can already notice that, using this procedure, one relies uniquely on considerations of analyticity and \textit{needs not} introduce any particular parametrization of the group element $g(\sigma,\tau)$ for the fluctuations around the classical solution, contrary to what is usually done in this type of analysis \cite{Frolov:2003qc,Frolov:2003tu,Arutyunov:2003za,Park:2005ji}. It is also nice to see that the fermionic and bosonic frequencies appear, in our approach, on a completely equal footing, both corresponding to simple poles which differ only by the sheets they unite - see figs.\ref{fig:bp} and \ref{fig:fp}. Finally, in principle, we can apply our method to any classical solution whereas the same generalization seems to be highly non-trivial to do directly in the action since we no longer have a simple field redefinition to make it time and space independent as was the case in \cite{Frolov:2003tu,Park:2005ji}.

\subsection{The BMN string } \la{BMN}
We shall consider the simplest possible solution amongst the family of circular strings presented in section \ref{circular}, the rotating point like BMN string \cite{Berenstein:2003gb} moving around a big circle of $S^5$. For this solution all spins except for
$$
\J_1=\J
$$
are set to zero. Then we have
$
m_1=0 \,\, , \,\, w_1=\J \,\, , \,\, \E=\kappa=\J \,.
$
For this solution the connection $A(x)$ presented in Appendix A is not only constant but also diagonal so we immediately find
\beq
\t p_{1,2} =- \t p_{3,4}  = \h p_{1,2} = -\h p_{3,4} =\frac{2\pi \J x}{x^2-1} \,. \label{pBMN}
\eeq
We see that this is indeed
the simplest $8$ sheet algebraic curve we could have built --  it has neither poles nor cuts connecting its sheets other than the trivial ones at $x=\pm 1$ (\ref{pm1}).

We shall now study the quantum fluctuations around this solution. For the sake of clarity we shall not write explicitly many of the quantities computed in the intermediate steps -- they can be found in Appendix B.

To consider the $16$ types of physical excitations we add all types of poles on the fig \ref{fig:bp} and \ref{fig:fp}. From (\ref{eqsposition}) we find that the poles in the physical domain with $|x|>1$, for this simple case, are all located at the same position
\beq
x_n^{ij}=x_n=\frac{1}{n}\(\J+\sqrt{\J^2+n^2}\) \,. \la{pos}
\eeq
Now we must find the quasi-momenta $p(x)+\delta p(x)$
\begin{itemize}
\item{with poles located at (\ref{pos}) with residues \eq{xnres} connecting the several sheets, }
\item{obeying the $x\rightarrow 1/x$ symmetry property \eq{x1/x},}
\item{with residues $\pm 1$ grouped as in (\ref{pm1}),}
\item{with large $x$ behavior given by (\ref{largex}).}
\end{itemize}

From the requirements listed above one can easily write the expression for the quasi-momenta. For example
\beqa
\la{BMNph2} \delta \hat p_2=\hat a+\frac{\delta\alpha_+}{x-1}+\frac{\delta\alpha_-}{x+1}
+\sum_{i=\h 3,\h 4,\t 3,\t 4}\sum_{n}\frac{\alpha(x_n^{\hat 2 i})N_n^{\hat 2 i}}{x-x_n^{\hat 2 i}}
-\sum_{i=\h 3,\h 4,\t 3,\t 4}\sum_{n}\frac{\alpha(x_n^{\hat 1 i})N_n^{\hat 1 i}}{1/x-x_n^{\hat 1 i}}  \\
\la{BMNph3}\delta \hat p_3=\hat b+\frac{\delta\beta_+}{x-1}+\frac{\delta\beta_-}{x+1}
-\sum_{i=\h 1,\h 2,\t 1,\t 2}\sum_{n}\frac{\alpha(x_n^{\hat 3 i})N_n^{\hat 3 i}}{x-x_n^{\hat 3 i}}
+\sum_{i=\h 1,\h 2,\t 1,\t 2}\sum_{n}\frac{\alpha(x_n^{\hat 4 i})N_n^{\hat 4 i}}{1/x-x_n^{\hat 4 i}}
\eeqa
where $\hat a, \hat b$ and $\delta\alpha_{\pm},\delta\beta_{\pm}$ are constants to be fixed and the last terms ensure the right poles in physical domain for $\delta \hat p_{1,4}(x)=-\delta \hat p_{2,3}(1/x)$. Similar expressions can be immediately written down for $\delta \hat p_{2,3}$ with the introduction of two new constants $\t a$ and $\t b$.

At this point we are left with the problem of fixing the eight constants
\beqa
\hat a , \hat b , \t a , \t b , \delta\alpha_{+},\delta\alpha_{-}, \delta\beta_{+},\delta\beta_{-}\,. \nn
\eeqa
This is precisely the number of conditions one obtains by imposing the $1/x$ behavior at large $x$ for the quasi-momenta (\ref{largex}) . The asymptotic of $\hat p_2,\hat p_3,\t p_2,\t p_3$ fix the first four constants while
the remaining four equations,  solvable only  if the level matching condition (\ref{hir}) is satisfied,
fix the remaining coefficients and yield
\beq
\delta E=\sum_{\rm All}\sum_n\frac{\sqrt{n^2+\J^2}-\J}{\J}N_n^{ij}+\sum_{{\rm AdS}^5}N^{ij}+\frac{1}{2}\sum_{{\rm Ferm}}N^{ij} \la{EBMN}
\eeq
where we indeed recognize the famous BMN frequencies \cite{Berenstein:2003gb} in the anomalous part of the energy shift.

\subsection{The circular string in $S^3$, the $1$ cut $su(2)$ solution}\label{1cut}
The next less trivial example is the simple $su(2)$ rigid circular string \cite{Frolov:2003qc}. Still it is simple enough so that
all results are explicit. This solution is obtained from the family of circular strings in section \ref{circular} by setting
\beq
m_1=-m_2=m \,\, , \,\, \J_1=\J_2=\J \nn
\eeq
with all other spins set to zero. For this solution
\beq
\E=\kappa=\sqrt{\J^2+m^2} \,. \nn
\eeq
The quasi-momenta can be computed as explained in Appendix A. The $AdS_5$ quasi-momenta are obtained as for the BMN string
\beq
 \h p_{1,2} = -\h p_{3,4} =\frac{2\pi \kappa \,x}{x^2-1}  \label{psu2A}
\eeq
while for the $S^5$ components $\t p_i$ we find that
this solution corresponds to 1 cut between $\t p_2$ and $\t p_3$ with mode number $k=-2m$, given by \cite{KMMZ}
\beqa
\(\bea{c}
\t p_1\\
\t p_2\\
\t p_3\\
\t p_4\\
\eea\)=
2\pi\(\bea{l}
+\frac{x}{x^2-1}K(1/x)\\
 +\frac{ x}{x^2-1}K(x)- m\\
 -\frac{ x}{x^2-1}K(x)+ m\\
-\frac{ x}{x^2-1}K(1/x)\\
\eea\)\, , \,\,\,\,\, K(x)\equiv \sqrt{m^2 x^2+\J^2}. \label{p_su(2)}
\eeqa
where we assume that $m>0$ and branch cut goes to the left of $x=-1$ so that
\beqa
&&K(x)= m x+\O(1/x)\,\, , \,\,K(x)=\J+\O(x)\nn \\
&&K(1)=K(-1)=\kappa>0 \nn
\eeqa

In the rest of this section we will compute the quantum
spectrum of the low lying excitations around this solution.
For simplicity we will consider the $AdS_5$, $S^5$ and fermionic fluctuations independently
assuming the level matching condition \eq{hir} to be satisfied \textit{for each of the sectors separately}. The result we give, however, is valid under the softer constraint (\ref{hir}) for all sectors, as one can easily check.

\subsubsection{Method of computation}\la{method}

Suppose we want to compute the variation
of the quasi-momenta $\delta p(x)$ when a small pole is added to some general finite gap solution with some square root cuts.  Since the branch points will be slightly displaced we conclude that $\delta p(x)$ behaves like $\d_{x_0}\sqrt{x-x_0}\sim 1/\sqrt{x-x_0}$ near each such point.

We are dealing with a 1-cut finite gap solution. Then, for $\delta\t p_2$, we can assume the most general analytical function with one branch cut, namely $f(x)+g(x)/K(x)$ where $f$ and $g$
are some rational functions and $K(x)$ was defined in (\ref{p_su(2)}). To obtain $\delta \t p_3$ it suffices to notice that (\ref{problem}) is simply telling us that $\delta \t p_3$ is the analytical continuation of $\delta \t p_2$ trough the cut. The remaining quasi-momentum $\delta \t p_{1,4}$ can then be obtained from this ones by the inversion symmetry (\ref{x1/x}). We conclude that
\beqa
 \(\bea{c}
\delta\t p_1\\
\delta\t p_2\\
\delta\t p_3\\
\delta\t p_4\\
\eea\)=
\(\bea{c}
-f(1/x)-\frac{g(1/x)}{K(1/x)}\\
 f(x)+\frac{g(x)}{K(x)}\\
 f(x)-\frac{g(x)}{K(x)}\\
-f(1/x)+\frac{g(1/x)}{K(1/x)}
\eea\) \label{fg} \,.
\eeqa

The only singularity of $\delta \tilde p_2$ apart from the branch cut are eventual simple poles at $\pm 1$ and $x_n$ and so the same must be true for $f(x)$ and $g(x)$. Then, just like in the previous example, these functions are uniquely fixed by the large
$x$ asymptotics \eq{largex} and by the residues at $x_n$ \eq{xnres} of the quasi-momenta.

Finally, since the $AdS_5$ part of the quasi-momenta of the non-perturbed finite gap solution has no branch cuts their variations $\delta \hat p_i$ have the same form \eq{BMNph2},\eq{BMNph3} as for simplest BMN string.

\subsubsection{The $AdS_5$ excitations}
This part is the simplest. The excitations live in the empty $AdS_5$ sheets where the only impact of the
$S^5$ classical solution comes through the Virasoro constraints, by the residues at $\pm 1$ (\ref{pm1}).
Thus the $\tilde p$ are nonperturbed
and $\delta\h p_i$ are the same as in BMN case \eqs{BMNph2}{BMNph3}
with only $AdS_5$ filling fractions $N$'s being nonzero. Indeed, comparing (\ref{pBMN}) and (\ref{psu2A}) we see that we can completely recycle the previous computation provided we replace $\J$ by $\kappa$ in the expression (\ref{pos}) for the pole's position. This leads immediately to
\beq
\delta E=\sum_{{\rm AdS}^5}\sum_n\frac{\sqrt{\J^2+m^2+n^2}}{\sqrt{\J^2+m^2}}N_n^{ij} \la{su2AdS}
\eeq

\subsubsection{The $S^5$ excitations}

We must now analyze the shift in quasi-momenta due to the excitation of the algebraic curve by the  four type of poles $(\t1\t3,\t2\t4,\t2\t3,\t1\t4)$. Since the
$AdS$ quasi-momenta are trivial, with no cuts, we obtain for $\delta \h p$ the same kind of expression we had for the BMN string (\ref{pBMN}), that is
\beq
\delta \h p_{1,2}=-\delta \h p_{3,4}= \frac{2 \pi \delta E}{\sqrt{\lambda}}\frac{x}{x^2-1} \,, \la{p12}
\eeq
where the constant factor was fixed by the asymptotics (\ref{largex})
\beq
\delta \h p_{1,2}\simeq-\delta \h p_{3,4} \simeq \frac{2 \pi \delta E}{\sqrt{\lambda}}\frac{1}{x} \,. \nn
\eeq
Due to the Virasoro constraints the poles at $\pm 1$ in the $AdS_5$ and $S^5$ sectors are synchronized (\ref{pm1}) so that we merely need to compute $f(x)$ and $g(x)$ from the large $x$ asymptotics \eq{largex} and the residue condition \eq{xnres} and extract, from these two functions, the residues at $\pm 1$. This is done in Appendix C.1. Let us just provide a glimpse of reasoning involved. Since the difference
$$
\delta \t p_3=f(x)-g(x)/K(x)
$$
must have a single pole at $x_n^{\t 1\t 3}$ with residue $\alpha(x_n^{\t 1\t 3})$ whereas the sum
$$
\delta \t p_2=f(x)+g(x)/K(x)
$$
must be analytical, we can, in this way, read the residues of both $f$ and $g$ at this point. This kind reasoning should be carried over for all the other excitations and for the points $x=\pm 1$ and leads to the ansatz (\ref{fsu2s5},\ref{gsu2s5}) where the only $3$ constants left to be found can be fixed by the large $x$ asymptotics of the quasi-momenta.

One can then read of the energy shift from the large $x$ asymptotics of the quasi-momenta
\beqa
\delta E=\sum_n (N_n^{\t 1\t 3}+N_n^{\t 2\t 4})\,\frac{x_n^{\t 1\t3}(m+n)-\J-K(x_n^{\t 1\t 3})}{\kappa}  +N_n^{\t1\t4}\,\frac{n\,x_n^{\t1\t4}-2\J}{\kappa}+N_n^{\t2\t3}\,\frac{2m+n}{x_n^{\t2\t3}\kappa} \la{su2S}
\eeqa
in terms of the positions of the roots obtained from the original algebraic curve through (\ref{eqsposition}).

\subsubsection{Fermionic excitations}
We have $8$ fermionic excitations but since
$\hat p_1=\hat p_2=-\hat p_3=-\hat p_4$ and $\tilde p_{1,2}=-\tilde p_{4,3}$
we will get the same result for the $(\hat 1\t 3,\hat 2\t 3,\hat 3\t 2,\hat 4\t 2)$ poles
and possibly another result for the $(\hat 1\t 4,\hat 2\t 4,\hat 3\t 1,\hat 4\t 1)$ excitations. We can repeat the same kind of calculations we did for the $S^5$ excitations to fix completely the quasi-momenta -- see Appendix C.2. Then, from the asymptotics (\ref{largex}) we get
\beqa
\delta \Delta
=\sum_{n}\(N_n^{\h1\t3}+N_n^{\h2\t3}+N_n^{\h3\t2}+N_n^{\h4\t2}\) \frac{m+n}{x_n^{\h1\t3}\kappa}
+\(N_n^{\h1\t4}+N_n^{\h2\t4}+N_n^{\h3\t1}+N_n^{\h4\t1}\)\frac{n\,x_n^{\h1\t4}-\J-\kappa}{\kappa} \,. \la{su2F}
\eeqa

\subsection{The circular string in $AdS_3$, the 1 cut $sl(2)$ solution}\la{sl2sol}

In section \ref{1cut} we analyzed in detail a simple $su(2)$ solution with a particular mode number $k=-2m$. In Appendix D we repeat the analysis for the general $sl(2)$ circular string \cite{Arutyunov:2003za,Park:2005ji} which also corresponds to a $1$--cut algebraic curve but this time with an arbitrary mode number $k$ for the cut \cite{Kazakov:2004nh}. This solution is again contained in the family of circular strings written in section \ref{circular}. It corresponds to two non zero spins
$$
\S_1=\S\,\, , \,\, \J_1=\J
$$
with mode numbers
$
m_1=m$, $k_1=k $
constrained by the level matching condition
$$
\S k+\J m=0
$$
and frequencies
$w_1=\J$ and ${\rm w}_1=w$
fixed by
\beq
w^3-(k^2+m^2+\J^2)w+2k m\J=0\la{eqw} \,.
\eeq
For this solution $\kappa=\sqrt{w^2-k^2}$ and the energy can be found from
\beq
\E=\kappa\(1+\frac{\S}{w}\) \nn
\eeq

In Appendix D we present the quasi-momenta associated to this classical solution and compute the fluctuation frequencies as we did for the $su(2)$ string. These results, together with the ones for the $su(2)$ circular string, are summarized and discussed in the following sections.

\section{Results, interpretation and $1$-loop shift}
\subsection{Results}
In this section we list all our results and introduce the notation usually used in the literature. In the next section we shall analyze them, compare them and draw some conclusions.

\subsubsection{Simple $su(2)$ circular string}

In section \ref{1cut} we found the level spacings around the simple $su(2)$ circular solution, that is the fluctuation frequencies of the effective quadratic Lagrangian obtained by expanding the Metsaev-Tseytlin action (\ref{ActionJ}) around this classical solution. In \cite{Frolov:2003tu} this computation was performed having in mind the stability analysis and computation of the one loop shift.
The various frequencies and corresponding degeneracies and origin can be summarized in table 1\footnote{By expanding the GS action without imposing the Virasoro conditions from the beginning, one obtains, apart from the frequencies listed in the above table, some massless modes with $\omega=n$ \cite{Frolov:2003qc}. This Virasoro modes can be seen from the Bethe ansatz point of view \cite{Gromov:2007fn} if an extra level of particles with rapidities $\theta$ is introduced \cite{GKSV}}.
\begin{table}[th]
\caption{\small\textsf{\textit{Simple $su(2)$ frequencies}}}
\beq \nn \bea{l|l|l}\toprule &\rm\bf\quad\quad\quad eigenmodes& \rm \bf notation
\\ \midrule
\bf\ S^5 &
    \bea{l}
    \sqrt{2\J^2+n^2\pm 2\sqrt{\J^4+n^2\J^2+m^2n^2}}\\
    \sqrt{\J^2+n^2-m^2} \\
    \eea &
    \bea{l}
    \omega_n^{S_\pm}\\
    \omega_n^{S}\\
    \eea
\\ \midrule
{\rm\bf Fermions} &
     \sqrt{\J^2+n^2}  &
    \bea{l}
    \omega_n^{F}
    \eea
\\ \midrule
\bf AdS_5 &
     \sqrt{\J^2+n^2+m^2}&
    \bea{l}
    \omega_n^{A}
    \eea
\\ \bottomrule
\eea \eeq
\end{table}
Using the notation introduced in this table we can replace the explicit expressions for the
position of the roots found from (\ref{eqsposition}) and recast our
results (\ref{su2AdS},\ref{su2S},\ref{su2F}) as
\beqa
\nn\kappa\ \delta E&=&\sum_n\(N_n^{\t1\t3}+N_n^{\t2\t4}\)\(\omega_{n+m}^S-\J\)+N_n^{\t2\t3}\omega_{n+2m}^{S_-}+N_n^{\t1\t4}\(\omega_{n}^{S_+}-2\J\)\\
\nn&+&\sum_n\(N_n^{\h1\t4}+N_n^{\h2\t4}+N_n^{\h3\t1}+N_n^{\h4\t1}\)\(\omega_{n}^F-\J+\frac{\kappa}{2}\)\\
&+&\sum_n\(N_n^{\h1\t3}+N_n^{\h2\t3}+N_n^{\h3\t2}+N_n^{\h4\t2}\)\(\omega_{n+m}^F-\frac{\kappa}{2}\) \nn\\
&+&\sum_n\(N_n^{\h1\h3}+N_n^{\h1\h4}+N_n^{\h2\h3}+N_n^{\h2\h4}\)\omega_{n}^A \,. \la{su2final}
\eeqa
We notice the appearance of constant shifts and relabeling of the frequencies when compared to those in table 1. We shall discuss this point below.

\subsubsection{General $sl(2)$ circular string}
The same analysis can be carried over for the $sl(2)$ circular string. In \cite{Arutyunov:2003za,Park:2005ji}
 this frequencies were computed and the result can be summarized in table 2\footnote{The results in this table
are slightly simplified compared to
those usually presented in the literature, especially the fermionic frequencies.}
\begin{table}[h] \la{t2}
\caption{\small\textsf{\textit{General $sl(2)$ frequencies}}}
\beq \nn \bea{l|l|l}\toprule \rm\bf &\rm\bf\quad\quad\quad eigenmodes& \rm \bf notation
\\ \midrule
 \bf AdS_5 &
    \bea{l}
     \(\omega^2-n^2\)^2+\frac{4\S}{{\rm w}} \kappa^2\omega^2-\frac{4\E}{\kappa}\(\omega {\rm w}-k n\)^2=0\\
      \sqrt{n^2+\kappa^2} \\
    \eea &
    \bea{l}
    \omega_n^{A_+}>\omega_n^{A_-}\\
    \omega_n^{A}\\
    \eea
\\ \midrule
 \text{\bf Fermions}  &
    \bea{l}
  \sqrt{\(n+\frac{\sqrt{w^2-\J^2}}{2}\)^2+\frac{1}{2}\(\kappa^2+\J^2-m^2\)}\\
    \eea
&
    \bea{l}
    \omega_n^{F}
    \eea
\\ \midrule
 \bf S^5 &
    \bea{l}
    \sqrt{\J^2+n^2-m^2}\\
    \eea
&
    \bea{l}
    \omega_n^{S}
    \eea
\\ \bottomrule
\eea \eeq
\end{table}

In the notation of the above table, the results \eq{sl2F},\eq{sl2S},\eq{E1324}\eq{E2314} derived in Appendix D can be put together as
\beqa
\nn\kappa\ \delta E&=&\sum_n\(N_n^{\h1\h3}+N_n^{\h2\h4}\)\omega_{n}^A
+N_n^{\h2\h3}\(\omega^{A_-}_{n-k}+w\)
+N_n^{\h1\h4}\(\omega_{n+k}^{A_+}-w\)\\
\nn&+&\sum_n\(N_n^{\h2\t3}+N_n^{\h2\t4}+N_n^{\h3\t1}+N_n^{\h3\t2}\)\(\omega_{n+m/2-k/2}^F-\omega_{m/2-k/2}^F+\frac{1}{2}\kappa\)\\
&+&\sum_n\(N_n^{\h1\t3}+N_n^{\h1\t4}+N_n^{\h4\t1}+N_n^{\h4\t2}\)\(\omega_{-n-m/2-k/2}^F-\omega_{-m/2-k/2}^F+\frac{1}{2}\kappa\) \nn \\
&+&\sum_n\(N_n^{\t1\t3}+N_n^{\t1\t4}+N_n^{\t2\t3}+N_n^{\t2\t4}\)\(\omega_{n+m}^S-\J\) \,. \la{Esl2}
\eeqa

\subsection{Explanation of shifts}

Let us first look at the $su(2)$ result (\ref{su2final}) and pick one if the frequencies, say the first one
\beq
\omega_{n+m}^S-\J \,. \label{example}
\eeq
We find two kinds of shifts relatively to the frequencies listed in the table 1, namely the constant shift $\J$ and the shift in the fourier mode $n\rightarrow n+m$. The same shifts we observe for the $sl(2)$ frequencies.

Let us understand the origin of this shifts. For that purpose consider a system of two harmonic oscillators,
\beqa
L_x=\frac{\dot{x}_1^2+\dot{x}_2^2}{2} -\frac{\omega^2}{2}\(x_1^2+x_2^2\) \,, \nn
\eeqa
and suppose that, instead of quantizing this system, we chose to quantize the system obtained by rotating $x_1,x_2$ with angular velocity $\J$, i.e. we move to the $y$ frame
\beq
x_1+ix_2=(y_1+iy_2) \,e^{i\J t} \,. \nn
\eeq
Then, we obtain\footnote{ In the $y$ frame the
 Lagrangian takes the form
$
2L_y=\dot{y}_1^2+\dot{y}_2^2-(\omega^2-\J^2)\(x_1^2+x_2^2\) +2\J y_1\dot y_2-2\J \dot y_1y_2 \,.
$}
\beqa
H_y= H_x+\J L_z \, , \nn
\eeqa
where $L_z$ is the usual angular momentum, so that
\beqa
E^y_{n_1,n_2}=\omega+\(\omega-\J\)n_1+\(\omega+\J\)n_2 \,.\nn
\eeqa
Thus for the radially symmetric wave function, for which $n_1=n_2$ (and in particular for the ground state energy), the constant shifts cancel and we obtain the same energies as for the first system. That, in general, the two results are different is obvious since the energy depends on the observer.

The constant shifts mentioned above have exactly this origin. In fact, when expanding the Metsaev-Tseytlin string action around the classical $su(2)$ circular string one obtains an effective \textit{time and space dependent} Lagrangian whose $\sigma,\tau$ dependence can be killed by a change of frame
\beq
\delta X=R(\sigma,\tau) \delta Y \nn
\eeq
where $\delta X$ are the (bosonic) components of the fluctuations and $R$ is a time and space dependent rotation matrix -- see for instance expression (2.14) in \cite{Frolov:2003tu}\footnote{The same is true for the $sl(2)$ circular string. The authors have moved to a different frame through a time and space dependent rotation -- see for instance equation 4.11 in \cite{Park:2005ji} -- and should, therefore, measure shifted energies.}. The same kind of field redefinitions are also present for the fermion fields. The time dependence of the rotation matrix gives the constant shifts as in the simple example we just considered while the space dependence in this change of frame is responsible for the relabeling of the mode numbers.

To make contact with the algebraic curve let us return to the
frequency \eq{example}
we picked as illustration. It corresponds to a pole
from sheet $\t p_1$ to $\t p_3$ (or from $\t p_2$ to $\t p_4$) whose
position is fixed by (\ref{eqsposition}). The result in the
rotated frame, $\omega^S_n$, would correspond to a pole with mode number $n+m$ whose position
is given by
\beq
\t p_1(x_n^{\tilde1\tilde 3})-\t p_3(x_n^{\tilde1\tilde 3})=2\pi n+2\pi m\,. \nn
\eeq
When plugging the actual expressions (\ref{p_su(2)}) for $\t p_1$
and $\t p_3$ in this equation we see that the $2 \pi m$ disappears and
the equation looks simpler than (\ref{eqsposition}). However, for several cut solutions there is no such obvious choice of mode numbers (or field redefinition
which kills the time dependence in the Lagrangian).

\subsection{$1$--loop shift and prescription for labeling fluctuation frequencies}
To compute the 1-loop shift to the classical energy of a given solution,
according to \cite{Frolov:2002av}, one has to sum over all energies of the modes in
the expansion around the classical configuration
\beq
\delta E_{1-\rm
loop}=\frac{1}{2\kappa}\lim_{N\rightarrow\infty}\sum\limits_{n=-N}^{N}
\(\sum\limits_{i=1}^{8}\Omega_{i,n}^B-\sum\limits_{i=1}^{8}\Omega_{i,n}^F\) \,. \nn
\eeq
However, the right hand side is hard to define rigorously. For $n\rightarrow\pm\infty$ each frequency behaves like
\beq
\Omega_{i,n}^{B}\simeq |n|\pm c^B_{i}+d^B_{i}\,\, , \,\,
\Omega_{i,n}^{F}\simeq |n|\pm c^F_{i}+d^F_{i}\,\, , \,\, \nn
\eeq
and thus the sum is sensible to the labeling of the frequencies. The seemingly innocent redefinition
\beq
\Omega_{i,n}^{B}\rightarrow \Omega_{i,{n+k_i}}^{B}\,\, , \,\,
\Omega_{i,n}^{F}\rightarrow \Omega_{i,{n+l_i}}^{F}\, ,\la{shifts}
\eeq
with integer shifts constrained by
$
\sum_i k_i-l_i=0
$
to ensure the convergence of the sum, \textit{does} change the result
\beq
\delta E_{1-\rm
loop}\rightarrow\delta E_{1-\rm
loop}+\sum (k^2_i-l_i^2+2 c_i^B k_i-2 c_i^F l_i) \, .\la{dife}
\eeq
In appendix E we discuss in greater detail the effect of these shifts.

One way to compute the frequencies is to expand the Metsaev-Tseytlin  action
around some classical solution. Generically, the resulting quadratic Lagrangian is
time and space dependent. To eliminate this dependence, when possible, a field
redefinition is performed. However, there are several ways to do the field redefinition to get a time and space independent action. Different choices will give different sets of frequencies related by transformations like \eq{shifts} and will therefore lead to different results. In appendix E we analyze this kind of dangers by focusing on two explicit examples.
 Thus we need a solid prescription for the labeling of the frequencies.

Suppose we were semi-classically quantizing around some classical
string solution in flat space. Then we would expect to find some
fluctuation frequencies, the zero modes, corresponding to an overall
translation of the string solution and which should, therefore, carry
no energy at all. Then the usual prescription is to take $\Omega_{i,0}=0$.

The zero modes should also exist for a string in the
$AdS_5\times S^5$ space with a large amount of isometries. Indeed, let
us take our results and denote the contribution at $n=0$ in \eq{su2final} and \eq{Esl2} by $\delta E_0^{su(2)}$ and $\delta E_0^{sl(2)}$. Then
we find that they are equal and given by
\beqa
\delta E_0 =\sum_{{\rm AdS}^5}N_{ij}+\frac{1}{2}\sum_{{\rm Ferm}}N_{ij}\,. \la{E0Delta}
\eeqa
In other words, the contribution to the anomalous part $\delta\Delta$
of zero modes for our labeling is zero! Thus the prescription we used seems to be the precise analogue of the flat space fourier modes prescription.

Moreover, by construction, we have a good BMN limit. That is, when in the limit of very small cuts with $m\rightarrow 0$ we recover the result \eq{EBMN} without any unusual shifts\footnote{This is not the case for the frequencies listed in table 2  for instance. For example, from this expression, we find, for the fermionic frequencies, $\omega^F_n\simeq \sqrt{\(n+k/2\)^2+\J^2}$. See also discussion in Appendix E.}.

Then, from (\ref{su2final}) and (\ref{Esl2}), we can write the $1$-loop shifts for the $1$-cut circular solutions\footnote{As for the simple example of the harmonic oscillators in the previous section, the sum of all constant shifts appearing in (\ref{su2final}) and (\ref{Esl2}) cancel so that only the shifts in mode number lead to a change of the final result. The difference with respect to the sum with no shifts can be obtained from \eq{dife} and is equal to $m^2/\kappa$ in both cases.}
\beqa
E_{1-loop}^{su(2)}\!\!\!&=&\!\!\!\frac{1}{2\kappa} \lim_{N\rightarrow\infty}\sum\limits_{n=-N}^{N} 4\,\omega^{A}_n+\omega^S_{n+m}+\omega^{S_-}_{n+2m}+\omega^{S_+}_n-4\,\omega^{F}_n-4\,\omega^{F}_{n+m} \nn \,,\\
E_{1-loop}^{sl(2)}\!\!\!&=&\!\!\!\frac{1}{2\kappa} \lim_{N\rightarrow\infty}\sum\limits_{n=-N}^{N} 2\,\omega_{n}^A +\omega^{A_+}_{n+k} +\omega_{n-k}^{A_-} +4\,\omega_{n+m}^S -4\,\omega_{n+\frac{m-k}{2}}^F
-4\,\omega_{-n-\frac{m+k}{2}}^F\,. \nn
\eeqa

\subsection{General $su(2)$ results}

Another interesting solution contained in the family of circular solutions described in section \ref{circular} is the generalization of the simple $su(2)$ solution to the case of
two non-equal spins $\J_{1,2}$ with two different mode numbers $m_{1,2}$. The fluctuation frequencies associated with this solution can be listed in table 3 \cite{Arutyunov:2003za}\footnote{The fermionic frequencies for the general circular string of section \ref{circular} can be computed (we shall publish our findings elsewhere). In particular, for the $su(2)$ general circular string we find the results listed in table 3.}

\begin{table}[h]
\caption{\small\textsf{\textit{General $su(2)$ frequencies}}}
\beq \nn \bea{l|l|l}\toprule \rm\bf &\rm\bf\quad\quad\quad eigenmodes& \rm \bf notation
\\ \midrule
 \bf S^5 &
    \bea{l}
     \(\omega^2-n^2\)^2 - \frac{4\J_2}{w_2}\(\omega w_1-m_1 n\)^2-\frac{4\J_1}{w_1}\(\omega w_2-m_2 n\)^2=0\\
      \sqrt{n^2+\nu^2} \\
    \eea &
    \bea{l}
    \omega_n^{S_+}>\omega_n^{S_-}\\
    \omega_n^{S}\\
    \eea
\\ \midrule
\rm\bf Fermions  &
    \bea{l}
  \sqrt{\(n-\frac{\sqrt{w_1^2+m_2^2-\kappa^2}}{2}\)^2+\J_1w_1+\J_2 w_2}\\
    \eea
&
    \bea{l}
    \omega_n^{F}
    \eea
\\ \midrule
 \bf AdS_5 &
    \bea{l}
    \sqrt{n^2+\kappa^2}\\
    \eea
&
    \bea{l}
    \omega_n^{A}
    \eea
\\ \bottomrule
\eea \eeq
\end{table}

Now, armed with our prescription, we can write the one loop shift unambiguously. Imposing
\begin{itemize}
\item Good BMN limit \eq{EBMN} for vanishing filling fractions,
\item Proper zero mode behavior with $n=0$ frequencies having trivial anomalous,
part \eq{E0Delta}.
\item For $m_1=-m_2=m$ we should retrieve the simple $su(2)$ result \eq{su2final},
\end{itemize}
we get (for $m_1+m_2\leq 0$)
\beqa
\nn\kappa\,\delta
E&=&\sum_n\(N_n^{\t1\t3}+N_n^{\t2\t4}\)\(\omega_{n+m_1}^S-w_1\)
+N_n^{\t2\t3}\(\omega_{n+m1-m2}^{S_-}+w_2-w_1\)\\
\nn&+&\sum_n N_n^{\t1\t4}\(\omega_{n+m_1+m_2}^{S_+}-w_2-w_1\)
+\sum_n\(N_n^{\h1\h3}+N_n^{\h1\h4}+N_n^{\h2\h3}+N_n^{\h2\h4}\)\omega_{n}^A\\
\nn&+&\sum_n\(N_n^{\h1\t4}+N_n^{\h2\t4}+N_n^{\h3\t1}+N_n^{\h4\t1}\)\(\omega_{n+\frac{m_1+m_2}{2}}^F-\omega_{\frac{m_1+m_2}{2}}^F+\frac{\kappa}{2}\)\\
&+&\sum_n\(N_n^{\h1\t3}+N_n^{\h2\t3}+N_n^{\h3\t2}+N_n^{\h4\t2}\)
\(\omega_{-n-\frac{m_1-m_2}{2}}^F-\omega_{-\frac{m_1-m_2}{2}}^F+\frac{\kappa}{2}\)
\,. \nn
\eeqa

\section{Conclusions}

We explained how to compute the quantum fluctuations around \textit{any} classical superstring motion in $AdS_5\times S^5$. These excitations include the fermionic, $AdS_5$ and $S^5$ modes. We showed that each mode correspond to adding a pole to a specific pair of sheets $i,j$ of the algebraic curve. The position of the pole is determined from the equation
\beq
p_{i}(x_n^{ i j})-p_{j}(x_n^{ i j})=2\pi n, \nn
 \eeq
and thus provides one with an \textit{unambiguous} labeling for the frequencies.
In particular we observe the nice feature that for $n=0$ this equation prescribes the pole at infinity, that is we find the expected zero modes associated with global transformations under the isometries of the target superspace.

Technically we computed the change in quasi-momenta due to the addition of these new poles and read, from the large $x$ asymptotics, the global charge corresponding to the $AdS$ Energy, that is the frequencies. However, since we computed explicitly the perturbed quasi-momenta we have obtained not only the energy shift but actually all conserved charges!

Finally our method gives a new insight into the 1-loop shift analysis.
Having at hand a simple and universal framework to compute fluctuation frequencies, we can now prove general statements about the 1-loop shift of \textit{any} classical string solution \cite{GV}.

\subsection*{Acknowledgements}
We would like to thank J.~Penedones, P.~Ribeiro, K.~Sakai, A.~Tseytlin, D.~Volin, K.~Zarembo and especially V.~Kazakov for many useful discussions.  The work of
N.G. was partially supported by French Government PhD fellowship,
by RSGSS-1124.2003.2 and by RFFI project grant 06-02-16786. P.~V.
is funded by the Funda\c{c}\~ao para a Ci\^encia e Tecnologia fellowship
{SFRH/BD/17959/2004/0WA9}.

\addtocontents{toc}{\protect\contentsline{section}{Appendix A: Quasi-momenta for a generic rigid circular string}{\arabic{page}\protect}}
\section*{Appendix A: Quasi-momenta for a generic rigid circular string}
To establish the link between the embedding coordinates solution (\ref{43}) with the coset's notations we introduce the matrices
\beqa
\mathcal{R} = \prod_{i=1}^{3} e^{\frac{i}{2} \(w_i\tau+i m_i \sigma\)\Phi_i}\cdot \mathcal{R}_0 \,\, \in SU(4) \nn
\eeqa
and
\beqa
\mathcal{Q} = e^{\frac{i}{2} \kappa\tau \Phi_1  }\cdot \prod_{i=1}^2  e^{-\frac{i}{2} \({\rm w}_i\tau+i k_i \sigma\)\Phi_{i+1}}\cdot \mathcal{Q}_0 \, \,\, \in SU(2,2) \,. \nn
\eeqa
where $\Phi_i$ are the Cartan generators,
\beqa
\Phi_{1}=  {\rm diag} \,\(+,+,-,-\)   \,\, , \,\, \Phi_{2}=  {\rm diag} \,\(+,-,+,-\)  \,\, , \,\,
\Phi_{3}=  {\rm diag} \,\(-,+,+,-\)  \,,  \nn
\eeqa
and $\mathcal{R}_0=e^{\Phi_{42} \theta}e^{\Phi_{64} \gamma}$ and $\mathcal{Q}_0=e^{\Phi'_{42} \psi}e^{\Phi'_{64} \rho}$ are constant matrices with
\beqa
&&\(\cos\gamma,\sin\gamma\cos\theta,\sin\gamma\sin\theta\)=\(\sqrt{\frac{\J_1}{w_1}},\sqrt{\frac{\J_2}{w_2}} ,\sqrt{\frac{\J_3}{w_3}} \) \,\, , \nn \\
&& \(\cosh\rho,\sinh\rho\cos\psi,\sinh\rho\sin\psi\)=\(\sqrt{\frac{\E}{\kappa}} ,\sqrt{\frac{\S_1}{{\rm w}_1}},\sqrt{\frac{\S_2}{{\rm w}_2}} \) \nn
\eeqa
and $\Phi_{42},\Phi_{64},\Phi'_{42},\Phi'_{64}$ given respectively by
\beqa
\frac{1}{2}\( \begin{array}{cccc}
0 & -1 & 0 & 0 \\
1 & 0 & 0 & 0 \\
0 & 0 & 0 & -1 \\
0 & 0 & 1 & 0
\end{array}\) \, , \, \( \begin{array}{cccc}
0 & 0 & 0 & 0 \\
0 & 0 & -1 & 0 \\
0 & 1 & 0 & 0 \\
0 & 0 & 0 & 0
\end{array}\) \, , \,
\frac{1}{2}\( \begin{array}{cccc}
0 & -1 & 0 & 0 \\
1 & 0 & 0 & 0 \\
0 & 0 & 0 & 1 \\
0 & 0 & -1 & 0
\end{array}\) \, , \, \frac{1}{2}\( \begin{array}{cccc}
0 & 0 & 0 & 1 \\
0 & 0 & -1 & 0 \\
0 & -1 & 0 & 0 \\
1 & 0 & 0 & 0
\end{array}\) \,. \nn
\eeqa

Let us ignore for a moment the fact that, for a generic choice of mode numbers $m_i,k_j$, these matrices are not always periodic. Then,
to describe the circular solutions we can use, as representative $g \in PSU(2,2|4)$, the block diagonal matrix
\beqa
g=\(\bea{c|c}
\mathcal{Q} &\;\;\;0\;\;\;\\ \midrule
\;\;\;0\;\;\;& \mathcal{R}
\eea\) \la{g} \,,
\eeqa
which indeed leads to (\ref{43}) under the map (\ref{map}).

What is particular about this solution is that, as follows trivially from the form of the matrices $\mathcal{R}$ and  $\mathcal{Q}$, the current
$$
J=-g^{-1}dg\,,
$$
and therefore also the flat connection $A(x)$ in (\ref{Ax}), are constant matrices! Then the computation of (\ref{monodromy}) is trivial and the quasi-momenta $p(x)$ are simply obtained from the eigenvalues of $\frac{2\pi}{i}A(x)$.

Before going on let us comment on the subtle point ignored above -- the periodicity of the rotation matrices $\mathcal{R}$ (and $\mathcal{Q}$). For some integers $m_i$ we see that this matrix could become anti-periodic. This means that in principle we should use another representative, $\mathcal{R}^{periodic}$, for which we should still have (\ref{map}) but which should be periodic. However, if both $\mathcal{R}$ and $\mathcal{R}^{periodic}$ obey these equations this means that they are related by an anti-periodic $SP(4)$ gauge transformation. This means that for the purpose of computing the quasi-momenta $p(x)$ we can indeed always use the element (\ref{g}) provided we keep in mind that if $\mathcal{R}$ is antiperiodic we can recover the real quasi-momenta through
\beqa
\{e^{i\h p_1},e^{i\h p_2},e^{i\h p_3},e^{i\h p_4}|e^{i\t p_1},e^{i\t p_2},e^{i\t p_3},e^{i\t p_4} \}_{
\begin{array}{ll} \text{For the true} \\
\text{representative $\mathcal{R}^{periodic}$} \end{array}} \nn \\
=\{e^{i\h p_1},e^{i\h p_2},e^{i\h p_3},e^{i\h p_4}|-e^{i\t p_1},-e^{i\t p_2},-e^{i\t p_3},-e^{i\t p_4} \}_{
\begin{array}{ll} \text{Using the anti--periodic } \\
\text{$\mathcal{R}$ instead} \end{array}} \nn \,.
\eeqa
The same kind of statement hold for the $AdS$ element $\mathcal{Q}$.

The computation of the quasi-momenta is then straightforward. The $S^5$ components $\tilde p_i$ are given in terms of the eigenvalues\footnote{To each eigenvalues we might need to add a multiple of $\pi$ in such a way that its asymptotics become those prescribed in section \ref{curve}. If $\mathcal{R}$ is periodic this multiple should contain an even number of $\pi$'s whereas if it is anti-periodic, we should add $\pi n$ with $n$ odd to each quasi-momenta -- see discussion in the text.} of the symmetric matrix
\beq
\tilde A(x)=\pi\(
\begin{array}{cccc}
-\t a_+(1/x)  & \t b_+ & - \t c(1/x)           & \t d(x) \\
\t b_+ & \t a_+(x)  & \t d(1/x)           &  \t c(x) \\
- \t c(1/x) & \t d(1/x) & \t a_-(x)             & \t b_- \\
\t d(x) & \t c(x) & \t b_- & -\t a_-(1/x) \\
\end{array}
\) \la{tA}
\eeq
%
with
\begin{eqnarray*}
\t a_{\pm}(x)&=&\pm \t a(x) -m_3\cos\theta \\
\t a(x)   &=&-\frac{m_1-w_1 x+\(m_2-w_2 x\) \cos
   \theta+x \cos 2 \gamma  (-w_1+m_1 x+(w_2-m_2 x) \cos \theta
   )}{x^2-1} \\
\t b_{\pm}&=&(m_2\mp m_3)\cos\gamma\sin\theta \\
\t c(x)&=&\frac{(m_2+m_3)x^2-(m_2-m_3)-2w_3 x}{x^2-1}\sin\gamma\sin\theta \\
\t d(x)&=&\frac{-m_1+w_1 x+(m_2-w_2\, x) \cos \theta  }{x^2-1} \sin 2 \gamma
\end{eqnarray*}
while the $AdS$ quasi-momenta $\h p_i$ are the eigenvalues of
\beq
\h A(x)=\pi\(
\begin{array}{cccc}
-\h a_+(1/x)  & \h b_+ & - \h c(x)           & \h d(x) \\
\h b_+ & \h a_+(x)  & \h d(x)           &  \h c(x) \\
\h c(x) &-\h d(x) & \h a_-(x)             & \h b_- \\
-\h d(x) & -\h c(x) & \h b_- & -\h a_-(1/x) \\
\end{array}
\) \la{hA}
\eeq
with
\begin{eqnarray*}
\h a_{\pm}(x)&=&\pm \frac{2\pi\kappa-k_1\(x^2-1\)\cos\theta}{x^2-1}\cosh\rho +k_2\cos\psi \\
\h b_{\pm}&=&(k_2\cosh\rho \mp k_1)\sin\psi \\
\h c(x)&=&\frac{k_2\(x^2+1\)-2{\rm w}_2 x  }{x^2-1}  \sin \psi \sinh \rho \\
\h d(x)&=&\frac{k_1\(x^2+1\)-2{\rm w}_1x  }{x^2-1}  \cos \psi \sinh \rho
\end{eqnarray*}
For the simple $su(2)$ or $sl(2)$ solutions we have, amongst other conditions, $\theta=\psi=0$ which simplifies the computation drastically.

\addtocontents{toc}{\protect\contentsline{section}{Appendix B: BMN string, details }{\arabic{page}\protect}}
\section*{Appendix B: BMN string, details}
This appendix serves as a complement to section \ref{BMN}. The
quasi-momenta with the correct poles located at (\ref{pos}) and
residues given by (\ref{xnres}) is given by
\beqa
\delta \hat p_2=\hat a+\frac{\delta\alpha_+}{x-1}+\frac{\delta\alpha_-}{x+1}
+\sum_{i=\h 3,\h 4,\t 3,\t 4}\sum_{n}\frac{\alpha(x_n^{\hat 2 i})N_n^{\hat 2 i}}{x-x_n^{\hat 2 i}}
-\sum_{i=\h 3,\h 4,\t 3,\t 4}\sum_{n}\frac{\alpha(x_n^{\hat 1 i})N_n^{\hat 1 i}}{1/x-x_n^{\hat 1 i}}\nn\\
\delta \hat p_3=\hat b+\frac{\delta\beta_+}{x-1}+\frac{\delta\beta_-}{x+1}
-\sum_{i=\h 1,\h 2,\t 1,\t 2}\sum_{n}\frac{\alpha(x_n^{\hat 3 i})N_n^{\hat 3 i}}{x-x_n^{\hat 3 i}}
+\sum_{i=\h 1,\h 2,\t 1,\t 2}\sum_{n}\frac{\alpha(x_n^{\hat 4 i})N_n^{\hat 4 i}}{1/x-x_n^{\hat 4 i}} \nn
\eeqa
where the last term guaranties that $\delta \hat p_{1,4}(x)=-\delta \hat p_{2,3}(1/x)$ have the right poles with the appropriate residues in the physical domain. Analogously
\beqa
\delta \tilde p_2=\t a+\frac{\delta\alpha_+}{x-1}+\frac{\delta\alpha_-}{x+1}
-\sum_{i=\h 3,\h 4,\t 3,\t 4}\sum_{n}\frac{\alpha(x_n^{\t 2 i})N_n^{\t 2 i}}{x-x_n^{\t 2 i}}
+\sum_{i=\h 3,\h 4,\t 3,\t 4}\sum_{n}\frac{\alpha(x_n^{\t 1 i})N_n^{\t 1 i}}{1/x-x_n^{\t 1 i}} \la{BMNpt2}\\
\delta \tilde p_3=\t b+\frac{\delta\beta_+}{x-1}+\frac{\delta\beta_-}{x+1}
+\sum_{i=\h 1,\h 2,\t 1,\t 2}\sum_{n}\frac{\alpha(x_n^{\t 3 i})N_n^{\t 3 i}}{x-x_n^{\t 3 i}}
-\sum_{i=\h 1,\h 2,\t 1,\t 2}\sum_{n}\frac{\alpha(x_n^{\t 4 i})N_n^{\t 4 i}}{1/x-x_n^{\t 4 i} \la{BMNpt3} }
\eeqa
and  $\delta \t p_{1,4}(x)=-\delta \t p_{2,3}(1/x)$. From the large $x$ behavior of these quasi-momenta one obtains
\beqa
&&\h a=-\sum_n\frac{2\pi n}{\sqrt{\lambda\,}\J} \!\!\sum_{i=\h 3,\h 4,\t 3,\t 4}\!\!\!\!N_n^{\h 1 i},\;\;\;\;\;
\h b=+\sum_n\frac{2\pi n}{\sqrt{\lambda\,}\J}\!\!\sum_{i=\h 1,\h 2,\t 1,\t 2}\!\!\!\!N_n^{\h 4 i} \nn \,,\\
&&\t a=+\sum_n\frac{2\pi n}{\sqrt{\lambda\,}\J}\!\!\sum_{i=\h 3,\h 4,\t 3,\t 4}\!\!\!\!N_n^{\t 1 i},\;\;\;\;\;
\t b=-\sum_n\frac{2\pi n}{\sqrt{\lambda\,}\J}\!\!\sum_{i=\h 1,\h 2,\t 1,\t 2}\!\!\!\!N_n^{\t 4 i} \, ,\nn
\eeqa
the level matching condition (\ref{hir})  and
\beqa
&&\delta\alpha^+-\delta\alpha^-=-\sum_n\frac{2\pi n}{\sqrt{\lambda\,}\J}\sum_{i=\h 3,\h 4,\t 3,\t 4}\ \sum_{j=\hat 1,\hat 2}N_n^{i j} \,,\nn \\
&&\delta\beta^+-\delta\beta^-=-\sum_n\frac{2\pi n}{\sqrt{\lambda\,}\J}\sum_{i=\h 3,\h 4,\t 3,\t 4}\ \sum_{j=\t 3,\t 4}N_n^{i j}\ . \nn
\eeqa

\addtocontents{toc}{\protect\contentsline {section}{Appendix C: $SU(2)$ circular string, details}{\arabic{page}}}
\section*{Appendix C: $SU(2)$ circular string, details}
In this appendix we present the details of the calculations from section \ref{1cut} of the fluctuation frequencies around the $1$-cut $SU(2)$ solution.
\addtocontents{toc}{\protect\contentsline{subsection}{\protect\numberline{C.1} $S^5$ modes}{\arabic{page}}}
\subsection*{C.1 $S^5$ modes}

We start from the ansatz \eq{fg}.
We have four types of poles $(\t1\t3,\t2\t4,\t2\t3,\t1\t4)$. Thus
\beq
f(x)=\frac{1}{2}\(\delta \t p_2(x)+\delta \t p_3(x)\) \nn
\eeq
must have simple poles at $x_n^{\t 2\t 4}$, $x_n^{\t 1\t 3}$ with residues $\frac{1}{2}\alpha(x_n^{\t 1\t 3})$ and $-\frac{1}{2}\alpha(x_n^{\t 2\t 4})$ respectively (see fig.\ref{fig:bp} or (\ref{xnres})). The same holds for
\beq
f(1/x)=-\frac{1}{2}\(\delta \t p_4(x)+\delta \t p_1(x)\) \,. \nn
\eeq
Moreover, since the residues of $\delta \t p_i$ are connected to the $AdS$ quasi-momenta (\ref{pm1}) and these are given by (\ref{p12}) we conclude that $f(x)$ should be regular at $x=\pm 1$. Thus we obtain
\beqa
f(x)&=&-\sum_{n}\(
\frac{N_n^{\t 2\t 4}}{2}\[\frac{\alpha(x_n^{\t 2\t 4})}{x-x_n^{\t 2\t 4}}
+\frac{\alpha(x_n^{\t 2\t 4})}{x_n^{\t 2\t 4}(1-x x_n^{\t 2\t 4})}\]
-({\t 2\t4} \rightarrow \t 1\t 3)\) \la{fsu2s5}
\eeqa
Then
\beq
g(x)=\frac{K(x)}{2}\(\delta \t p_2(x)-\delta \t p_3(x)\) \nn
\eeq
must have simple poles at $x_n^{\t 2\t 4}$, $x_n^{\t 1\t 3}$ and  $x_n^{\t 2\t 3}$ with residues $-\frac{1}{2}\alpha(x_n^{\t 1\t 3})$, $-\frac{1}{2}\alpha(x_n^{\t 2\t 4})$ and $-\alpha(x_n^{\t 2\t 3})$ respectively while
\beq
g(1/x)=\frac{K(x)}{2}\(\delta \t p_4(x)-\delta \t p_1(x)\) \nn
\eeq
must have simple poles at $x_n^{\t 2\t 4}$, $x_n^{\t 1\t 3}$ and  $x_n^{\t 1\t 4}$ with residues $\frac{1}{2}\alpha(x_n^{\t 1\t 3})$, $\frac{1}{2}\alpha(x_n^{\t 2\t 4})$ and $\alpha(x_n^{\t 1\t 4})$ respectively. Contrary to $f(x)$, this function may have poles at $\pm 1$ so we arrive at
\beqa
\nn g(x)&=&a+\frac{\alpha_-}{x^2-1}+\frac{x\alpha_+}{x^2-1}+\sum_n\(
N_n^{\t1\t4}\frac{\alpha(x_n^{\t1\t4})K(1/x_n^{\t1\t4})}{x_n^{\t1\t4}(1-xx_n^{\t1\t4})}
-N_n^{\t2\t3}\frac{\alpha(x_n^{\t2\t3})K(x_n^{\t2\t3})}{x-x_n^{\t2\t3}}
\)\\
&+&\sum_n\(\frac{N_n^{\t 2\t 4}}{2}\[
\frac{\alpha(x_n^{\t 2\t4})K(1/x_n^{\t 2\t4})}{x_n^{\t 2\t4}(1-xx_n^{\t 2\t4})}-\frac{\alpha(x_n^{\t 2\t4})K(x_n^{\t 2\t4})}{x-x_n^{\t 2\t4}}\]+
({\t 2\t4} \rightarrow \t 1\t 3)
\)  \,. \la{gsu2s5}
\eeqa
Finally the remaining constants are fixed by the large $x$ asymptotic \eq{largex} to be
\beqa
a&=&-\frac{2\pi}{\sqrt\lambda}\sum_n\[m(N_n^{\t 1\t 3}+N_n^{\t 2\t 4})+2mN_n^{\t 2\t 3}\] \nn\\
\nn \alpha_+&=&\frac{2\pi}{\sqrt\lambda}\sum_n\[(N_n^{\t 1\t 3}+N_n^{\t 2\t 4})\(x_n^{\t 1\t3}(m+n)-\J-K(x_n^{\t 1\t 3})\) +N_n^{\t1\t4} (x_n^{\t1\t4}n-2\J)+N_n^{\t2\t3}\frac{2m+n}{x_n^{\t2\t3}}\]\\
\alpha_-&=&\frac{2\pi}{\sqrt\lambda}\sum_n\[N_n^{\t 1\t 3}+N_n^{\t 1\t 4}+N_n^{\t 2\t 3}+N_n^{\t 2\t 4}\]n  \nn \,.
\eeqa
Then, from the residue at $x=1$ we read $
\delta E=\frac{\alpha_+}{\sqrt{m^2+\J^2}} $.
\addtocontents{toc}{\protect\contentsline{subsection}{\protect\numberline{C.2} Fermionic modes}{\arabic{page}}}
\subsection*{C.2 Fermionic modes}
Arguments similar to the ones in the previous section lead to
\beqa
f(x)=\frac{2\pi}{\sqrt\lambda} \frac{x}{x^2-1}\sum_n\[\frac{N_n^{\h 1\t 4}+N_n^{\h 2\t 4}-N_n^{\h 3\t 1}-N_n^{\h 4\t 1}}{x x_n^{\h 1\t 4}-1}
+x\frac{N_n^{\h 3\t 2}+N_n^{\h 4\t 2}-N_n^{\h 1\t 3}-N_n^{\h 2\t 3}}{x - x_n^{\h 3\t 2}}\] \nn
\eeqa
and
\beqa
g(x)&=&b+\frac{\beta_-}{x^2-1}+\frac{x\beta_+}{x^2-1} \nn\\
\nn&+&\frac{2\pi}{\sqrt\lambda}\sum_n\[\frac{\(\sqrt{m^2+\J^2}x_n+n(1-x_n^2)\)(N_n^{\h 1\t 4}+N_n^{\h 2\t 4}+N_n^{\h 3\t 1}+N_n^{\h 4\t 1})}{(1-x x_n)(x_n^2-1)}\right.\nn\\
\nn&-&\left.x_n\frac{\(\sqrt{m^2+\J^2}x_n+(n+m)(1-x_n^2)\)(N_n^{\h 3\t 2}+N_n^{\h 4\t 2}+N_n^{\h 1\t 3}+N_n^{\h 2\t 3})}{(x - x_n)(x_n^2-1)}\]\nn
\eeqa
where
\beqa
b&=&\frac{2\pi m}{\sqrt\lambda}\sum_n(N_n^{\h1\t3}+N_n^{\h2\t3}+N_n^{\h1\t3}+N_n^{\h1\t3}) \nn\\
\nn \beta_-&=&\frac{2\pi}{\sqrt\lambda}\sum_n\(\!\frac{x_n\sqrt{m^2+\J^2}}{x_n^2-1}-n\!\)\!(N_n^{\h1\t3}+N_n^{\h1\t4}+N_n^{\h2\t3}+N_n^{\h2\t4}+
N_n^{\h3\t1}+N_n^{\h3\t2}+N_n^{\h4\t1}+N_n^{\h4\t2})\\
\nn \beta_+&=&\frac{2\pi}{\sqrt\lambda}\sum_n\[\(\J-n x_n+\frac{x_n^2\sqrt{m^2+\J^2}}{x_n^2-1}\)(N_n^{\h1\t4}+N_n^{\h2\t4}+N_n^{\h3\t1}+N_n^{\h4\t1})\right.\\
&+&\left.\(\frac{\sqrt{m^2+\J^2}}{x_n^2-1}-\frac{m+n}{x_n}\)
(N_n^{\h1\t3}+N_n^{\h2\t3}+N_n^{\h3\t2}+N_n^{\h4\t2}) \] \,.\nn
\eeqa
The $AdS_5$ part of the quasi-momenta is given by
\beqa
\delta \h p_2(x)=\frac{2\pi}{\sqrt\lambda}\frac{x}{x^2-1}\(+\delta\Delta-2\frac{N_n^{\h1\t3}+N_n^{\h1\t4}}{x x_n-1}-2x\frac{N_n^{\h2\t3}+N_n^{\h2\t4}}{x_n-x}\) \nn\\
\delta \h p_3(x)=\frac{2\pi}{\sqrt\lambda}\frac{x}{x^2-1}\(-\delta\Delta+2\frac{N_n^{\h4\t1}+N_n^{\h4\t2}}{x x_n-1}+2x\frac{N_n^{\h3\t1}+N_n^{\h3\t2}}{x_n-x}\) \nn
\eeqa
and $\delta\h p_{1,4}(x)=-\delta\h  p_{2,3}(1/x)$. The constant $\Delta$ can be found from fixing the residues at $\pm 1$ for $\delta \h p_i$ and $\delta \t p_i$ to be equal (\ref{pm1}) and is given in the main text \eq{su2F}.

\addtocontents{toc}{\protect\contentsline {section}{Appendix D: $SL(2)$ circular string}{\arabic{page}}}
\section*{Appendix D: $SL(2)$ circular string}
The eigenvalues of (\ref{tA}) and (\ref{hA}) for the $sl(2)$ circular string described in the beginning of section \ref{sl2sol} yield the most general $1$-cut quasi-momentum connecting $\h p_2$ and $\h p_3$ (Due to the $x\rightarrow 1/x$ symmetry, $\h p_1$ and $\h p_4$ will be also connected by a cut, but this will be in unphysical domain, that is inside the unit circle). Explicitly, we find
\beq
\t p_{1,2} =- \t p_{3,4}  = 2\pi \,\frac{ \J x+  m}{x^2-1} \,.\la{psl2S}
\eeq
and
\beqa
 \(\bea{c}
\h p_1\\
\h p_2\\
\h p_3\\
\h p_4\\
\eea\)=
 \(\bea{l}
-\h p_2(1/x)\\
+\h p_2(x)\\
-\h p_2(x)\\
+\h p_2(1/x)\\
\eea\) \,. \nn
\eeqa
where \cite{Kazakov:2004nh}
\beq
\h p_2=k\pi\(1-\frac{(C x+1)\sqrt{x^2-2BCx+C^2}}{C(x^2-1)}\) \,.\la{1cutsl2}
\eeq
and
\beq
w\equiv \frac{k}{2}\(C+\frac{1}{C}\) \,\, , \,\, B=1+\frac{2\S}{w}\la{EB}  \,.
\eeq
From all solutions of (\ref{eqw},\ref{EB}) for solutions for $C$ and $B$ we should pick the one for which we have a real cut outside the unit circle.

In the rest of this appendix we will excite this solution by adding poles
to quasi-momenta as we did for the $su(2)$ solution. In this way we shall find the energy shifts around this
classical solution. Moreover, as for the $su(2)$ string, we shall consider the
${\rm AdS}_5, {S}^5$ and fermions separately, assuming for simplicity
the Riemann identity \eq{hir} to be satisfied \textit{for each of the
sectors separately} -- the result, as before, holds if we relax this
stronger assumption.

\addtocontents{toc}{\protect\contentsline{subsection}{D.1 The $AdS_5$ excitations }{\arabic{page}\protect}}
\subsection*{D.1 The $AdS_5$ excitations}
For these excitations the $S^5$ quasi-momenta remains untouched because its asymptotics do not change and it is still only allowed to have simple poles at $\pm 1$. Due to the Virasoro coupling of these quasi-momenta to the $AdS_5$ ones through the poles at $\pm 1$ (\ref{pm1}), we see that $\delta \h p_i$ must have no poles at these points. The only poles of these quasi-momenta should be located at $x_n^{\h i\h j}$ with residues given by (\ref{xnres}) -- see fig.\ref{fig:bp}.
Finally, as explained in section \ref{method}, the perturbed quasi-momenta should have  inverse square behavior close to the branch points of the classical solution.

Thus, from the same kind of reasoning we saw in the previous section for the $su(2)$ circular string, we find (\ref{p2314}) for the $AdS_3$ excitations connecting sheets $(\h p_2,\h p_3)$ and $(\h p_1,\h p_4)$ and
(\ref{p1324}) for the remaining $AdS_5$ excitations uniting
$(\h p_1,\h p_3)$ and $(\h p_2,\h p_4)$. From the large $x$ behavior (\ref{largex}) of these quasi-momenta we read the energy shifts\footnote{these $AdS_3$ excitations were also found in a similar way by K.Zarembo in relation with the finite size corrections computation~\cite{Schafer-Nameki:2005tn} (according to the
private communication).}
\beq
\delta E=\sum_n \(N_n^{\h 2\h 3}\[\frac{k-n}{k}\frac{x_n^{\h2\h3}-C^{-1}}{x_n^{\h2\h3}+C^{-1}}+\frac{nw}{k\kappa}\]+N_n^{\h 1\h 4}\[\frac{k+n}{k}\frac{x_n^{\h1\h4}-C}{x_n^{\h1\h4}+C}+\frac{nw}{k\kappa}\]\) \label{E2314}
\eeq
and
\beq
\delta E=\sum_n  N_n^{\h 1\h 3}
\[\frac{K(x_n^{\h1\h3})k+n(x_n^{\h1\h3}-C)}{k(x_n^{\h1\h3}+C)}+\frac{n w}{k\kappa}\] +N_n^{\h 2\h 4}\[\frac{K(x_n^{\h2\h4})k+n(x_n^{\h2\h4}-C)}{k(x_n^{\h2\h4}+C)}+\frac{n w}{k\kappa}\] \label{E1324}
\eeq
where, as for the $su(2)$ string, we denote the square root in the classical solution (\ref{1cutsl2}) by $K(x)$.

\addtocontents{toc}{\protect\contentsline{subsubsection}{D.1.1 $AdS_3$ excitations -- details }{\arabic{page}\protect}}
\subsubsection*{D.1.1 $AdS_3$ excitations -- details}
In strict analogy to what we have already seen in for $SU(2)$ solution, the two (left and right) physical excitations inside the $SL(2)$ sector described by the $AdS_3\ \sigma-$model, are given by the poles, connecting the pairs of sheets $(\h p_2,\h p_3)$ and $(\h p_1,\h p_4)$.
The number of such poles we denote $N_{\h 2\h 3}$ and $N_{\h 1\h 4}$ respectively.

As explained above, the $AdS_5$ excitations shifts of $\h p$'s have no poles at $\pm 1$ and must present an inverse square root behavior close to the branch points of the classical solution. Thus we can write
\beq
\delta \h p_{2}(x)=\frac{2\pi}{\sqrt\lambda K(x)}\sum_{n}\(N_n^{\h 2\h 3}\frac{x a_n}{x-x_n^{\h 2\h 3}}+N_n^{\h 1\h 4}\frac{x \bar a_n}{x-1/x_n^{\h 1\h 4}}\) \la{p2314}
\eeq
where
$$
K(x)\equiv \sqrt{x^2-2BC\,x+C^2}
$$
and the position of the roots is given by (\ref{eqsposition}).
Fixing the residues at $x_n^{\h 2\h3}$ and $x_n^{\h 1\h4}$ according to (\ref{xnres}) we have
\beq
a_n=\frac{2 C x_n^{\h2 \h3}(k-n)}{k(C x_n^{\h2\h3}+1)},\;\;\;\;\;\bar a_n=-\frac{2 C(k+n)}{k(C+ x_n^{\h1\h4})} \nn
\eeq
The large $x$ asymptotic \eq{largex} is consistent if Riemann bilinear
identity \eq{hir} is satisfied
\beq
\sum_n \(N_n^{\h 2\h 3} + N_n^{\h 1\h 4}\)n=0 \nn
\eeq
and then the energy shift is given by (\ref{E2314}).

\addtocontents{toc}{\protect\contentsline{subsubsection}{D.1.2 The remaining $AdS_5$ excitations -- details}{\arabic{page}\protect}}
\subsubsection*{D.1.2 The remaining $AdS_5$ excitations -- details}
These correspond to simple poles connecting $(\h p_1,\h p_3)$ and $(\h
p_2,\h p_4)$ for which
\beq
\delta \h p_2(x)=\frac{2\pi}{\sqrt\lambda }\sum_n\[ \frac{N_n^{\h 1\h 3}x\(a_n+\frac{b_n+c_nx}{K(x)}\)}{(x-x_n^{\h 1\h3})(x-1/x_n^{\h 1\h3})}
+\frac{N_n^{\h 2\h 4}x\(\bar a_n+\frac{\bar b_n+\bar c_nx}{K(x)}\)}{(x-x_n^{\h 2\h4})(x-1/x_n^{\h 2\h4})}\] \,.\la{p1324}
\eeq
Then $\delta\h p_3$, just as we saw for the $su(2)$ solution, is the analytical continuation of $\delta \h p_2$ through the cut. In simpler terms, it corresponds to a simple change of sign of $K(x)$ in the above expression. Finally $\delta\h p_{1,4}(x)=-\delta\h p_{2,3}(1/x)$.

The undetermined coefficients are fixed by the residues\\[-0.9cm]
\beqa
\nn \res{x=x_n^{\h 1\h 3}}\h p_{1,3}=\pm\alpha\(x_n^{\h 1\h 3}\)N_n^{\h1\h3},\;\;\;\;\;\res{x=x_n^{\h 1\h 3}}\h p_{2,4}=0\\[-0.5cm]
\nn\res{x=x_n^{\h 2\h 4}}\h p_{2,4}=\pm\alpha\(x_n^{\h 2\h 4}\)N_n^{\h2\h4},\;\;\;\;\;\res{x=x_n^{\h 2\h 4}}\h p_{1,3}=0
\eeqa
to be
\beqa
\nn a_n=-1,\;\;\;\;\;b_n=C\frac{2n x_n^{\h 1\h3}+k K(x_n^{\h 1\h3})}{k(x_n^{\h 1\h 3}+C)},\;\;\;\;\;c_n=\frac{k K(x_n^{\h1\h3})-2Cn}{k(C+x_n^{\h1\h3})}\\
\nn\bar a_n=1,\;\;\;\;\;\bar b_n=C\frac{2n x_n^{\h 2\h4}+k K(x_n^{\h 2\h4})}{k(x_n^{\h 2\h 4}+C)},\;\;\;\;\;\bar c_n=\frac{k K(x_n^{\h2\h4})-2Cn}{k(C+x_n^{\h2\h4})}
\eeqa
Hence, with the level matching condition
\beq
\sum_n \(N_n^{\h 1\h 3} + N_n^{\h 2\h 4}\)n=0 \,,\nn
\eeq
we find, from the large $x$ behavior, the energy shift (\ref{E1324}).

\addtocontents{toc}{\protect\contentsline{subsection}{D.2 The $S^5$ excitations}{\arabic{page}\protect}}
\subsection*{D.2 The $S^5$ excitations}
The $S^5$ quasi-momentum (\ref{psl2S}) has no branch cuts and thus $\delta \t p_i$ will be of the same form as we found for the BMN string except that the position of the roots, found from (\ref{eqsposition}) is now given by
\beq
\t x_n\equiv x_n^{\t1\t3}=x_n^{\t1\t4}=x_n^{\t2\t3}=x_n^{\t1\t4}=\frac{\J+\sqrt{\J^2+(n+m)^2-m^2}}{n} \label{possl2}
\eeq
instead of (\ref{pos}). The explicit expressions for $\delta \t p_i$ are given in (\ref{ptS}). This perturbation shifts the residues at $\pm 1$ from
$$
\pi \(\J\mp m\)
$$
to some other values which we parameterize by
$$
\pi \(\J+\delta \J^{eff} \mp \(m + \delta m^{eff} \) \) \,.
$$
The precise expressions for these shifts can be found in (\ref{shift1},\ref{shift2}). But then, since the $AdS_5$ quasi-momenta only knows about the $S^5$ sector though the residues at these points, the perturbed quasi-momenta $\h p_i+\delta \h p_i$ will be given by the same expression (\ref{1cutsl2}) with the trivial replacement
$$\J,m\rightarrow \J+\delta \J^{eff},m+ \delta m^{eff} \,.$$ The same is true for the energy, given in (\ref{EB}) so that -- see Appendix D.2 for details -- we can immediately find
\beq
\delta E
=\frac{1}{\kappa}\sum_n\(N_n^{\t 1\t3}+N_n^{\t 1\t4}+N_n^{\t 2\t3}+N_n^{\t 2\t4}\)\(\sqrt{(n+m)^2-m^2+\J^2}-\J
\) \,. \label{sl2S}
\eeq

\addtocontents{toc}{\protect\contentsline{subsubsection}{D.2.1 The $S^5$ excitations -- details}{\arabic{page}\protect}}
\subsubsection*{D.2.1 The $S^5$ excitations -- details}
As explained above the perturbed $S^5$ quasi-momenta are of the BMN form (\ref{BMNpt2},\ref{BMNpt3})
\beqa
\delta \t p_2(x)=+\frac{4\pi}{\sqrt\lambda}\frac{x^2}{x^2-1}\sum_n
\(\frac{N^{\t2\t3}_n+N^{\t2\t4}_n}{\t x_n-x}+\frac{N^{\t1\t3}_n+N^{\t1\t4}_n}{x^2\t x_n-x}\) \nn \\
\nn\delta \t p_3(x)=-\frac{4\pi}{\sqrt\lambda}\frac{x^2}{x^2-1}\sum_n
\(\frac{N^{\t1\t3}_n+N^{\t2\t3}_n}{\t x_n-x}+\frac{N^{\t2\t4}_n+N^{\t1\t4}_n}{x^2\t x_n-x}\) \la{ptS}
\eeqa
where $\t x_n$ is given by (\ref{possl2}). Then, in the notation introduced above, the shift in the $x=\pm 1$ residues is given by
\beqa
\delta\J^{\rm eff}&=&\frac{\sum_n\(N_n^{\t 1\t 3}+N_n^{\t 1\t 4}+N_n^{\t 2\t 3}+N_n^{\t 2\t 4}\)\(mn+\J^2-\J\sqrt{\J^2+n^2+2mn}\)}{\sqrt\lambda(m^2-\J^2)} \la{shift1}
\eeqa
while $\delta m^{\rm eff}$ is given by
\beq
\J\delta m^{\rm eff}+\delta\J^{\rm eff} m=\frac{1}{\sqrt\lambda}\sum_n\(N^{\t 1\t 3}_n+N^{\t 1\t 4}_n+N^{\t 2\t 3}_n+N^{\t 2\t 4}_n\)n=0 \la{shift2} \,
\eeq
due to the Riemann condition. Then, from (\ref{EB},\ref{eqw}), we have
\beq
\delta\E=\frac{w^3-k m \J}{w^2\kappa}\delta w=\frac{w(k^2+m^2+\J^2)-3 k m \J}{w^2\kappa}\delta w \nn
\eeq
where, using (\ref{eqw},\ref{shift2}), we have
\beq
\delta w=-\frac{\delta \J^{eff}}{\J}\frac{w^2(m^2-\J^2)}{w(k^2+m^2+\J^2)-3km\J} \nn
\eeq
so that $
\delta E$ will be given by \eq{sl2S}

\addtocontents{toc}{\protect\contentsline{subsection}{D.3 Fermionic excitations}{\arabic{page}\protect}}
\subsection*{D.3 Fermionic excitations}
The fermionic excitations can be treated as for the $su(2)$ string. As before we expect at most two different answers for the energy shifts -- one coming from the poles uniting the $(\hat 1\t 3,\hat 1\t 4,\hat 4\t 1,\hat 4\t 2)$ sheets another result for the $(\hat 2\t 3,\hat 2\t 4,\hat 3\t 1,\hat 3\t 2)$. From the expressions in Appendix C.3 we find
\beqa
\delta \Delta
=\sum_{n}\(N_n^{\h1\t3}+N_n^{\h1\t4}+N_n^{\h4\t1}+N_n^{\h4\t2}\) \delta\Delta_n^{(1)} +\sum_{n}\(N_n^{\h2\t3}+N_n^{\h2\t4}+N_n^{\h3\t1}+N_n^{\h3\t2}\)\delta\Delta_n^{(2)} \,, \la{sl2F}
\eeqa
where
\beqa
\delta\Delta_n^{(1)}&=&\frac{(2m+k)-2\J C-k C^2}{k(C^2-1)(C^{-1}x_n^{\h 1\t 3}+1)}+\frac{n(C^{-1}x_n^{\h 1\t 3}-1)}{k(C^{-1}x_n^{\h 1\t 3}+1)}+\frac{nw}{k\kappa} \nn \\
\delta\Delta_n^{(2)}&=&\frac{(2m-k)C^2-2\J C+k}{k(C^2-1)(Cx_n^{\h 2\t
3}+1)}-\frac{n(Cx_n^{\h 2\t 3}-1)}{k(Cx_n^{\h 2\t 3}+1)}+\frac{nw}{k\kappa} \nn
\eeqa
with the position of the fermionic poles being given by (\ref{eqsposition}), in terms of the algebraic curve for the classical solution.

\addtocontents{toc}{\protect\contentsline{subsubsection}{D.3.1 Fermionic excitations -- details}{\arabic{page}\protect}}
\subsubsection*{D.3.1 Fermionic excitations -- details}
The $S^5$ part of the quasi-momenta is given by
\beqa
\t p_2(x)=+\frac{4\pi x}{\sqrt\lambda(x^2-1)}\sum_n\(\frac{N_n^{\h 3\t 1}+N_n^{\h 4\t 1}}{x x_n^{\h 3\t 1}-1}-x\frac{N_n^{\h 3\t 2}+N_n^{\h 4\t 2}}{x- x_n^{\h 3\t 1}}\)\nn \\
\t p_3(x)=-\frac{4\pi x}{\sqrt\lambda(x^2-1)}\sum_n\(\frac{N_n^{\h 1\t 4}+N_n^{\h 2\t 4}}{x x_n^{\h 3\t 1}-1}-x\frac{N_n^{\h 1\t 3}+N_n^{\h 2\t 3}}{x- x_n^{\h 3\t 1}}\) \nn
\eeqa
whereas the $AdS_5$ part is more complicated. Parameterizing $\delta\h p$ as we did for the $su(2)$ string in (\ref{fg}), we have
\beqa
\nn f(x)&=&\frac{x}{x^2-1}\sum_n\(x\frac{N_n^{\h2\t3}+N_n^{\h2\t4}-N_n^{\h3\t1}-N_n^{\h3\t2}}{x-x_n^{\h2\t3}}
+\frac{N_n^{\h2\t3}+N_n^{\h2\t4}-N_n^{\h3\t1}-N_n^{\h3\t2}}{xx_n^{\h2\t3}-1}\)\\
 \nn g(x)&=&\frac{x}{x^2-1}\sum_n\(\[\frac{x K(x_n^{\h2\t3})}{x-x_n^{\h2\t3}}+a_n x+b_n\](N_n^{\h2\t3}+N_n^{\h2\t4}+N_n^{\h3\t1}+N_n^{\h3\t2})\right.\\
\nn&&\hspace{1.7cm}-\left.\[\frac{x x_n^{\h1\t3} K(1/x_n^{\h1\t3})}{xx_n^{\h1\t3}-1}+\bar a_n x+\bar b_n\]\(N_n^{\h1\t3}+N_n^{\h1\t4}+N_n^{\h4\t1}+N_n^{\h4\t2}\)
\)
\eeqa
where the remain constants are given by
\beqa
\nn a_n&=&C\frac{(C^2-1)(k-2n)x_n^{\h2\t3}+2C m-2\J}{(C^2-1)(C x_n^{\h2\t3}+1)k}\\
\nn \bar a_n&=&C\frac{(C^2-1)k-2(m+n)+2C(Cn+\J)}{(C^2-1)(C +x_n^{\h1\t3})k}\\
\nn  b_n&=&C\frac{(C^2-1)k-2C^2(m+n)+2(n+C\J)}{(C^2-1)(C +x_n^{\h2\t3})k}\\
\nn \bar b_n&=&-C\frac{(C^2-1)(k +2n)x_n^{\h1\t3} +2C m-2C^2\J}{(C^2-1)(C +x_n^{\h1\t3})k}
\eeqa
Then the energy shifts can be read from the large $x$ asymptotics and are given in (\ref{sl2F}).

\addtocontents{toc}{\protect\contentsline{section}{Appendix E: Ambiguities due to shifts}{\arabic{page}\protect}}
\section*{Appendix E: Ambiguities due to shifts}
To compute the 1-loop shift one must sum all frequencies. This sum, however,
is sensitive to the way the frequencies are labeled. Let us
demonstrate this on a simple example. Consider the sum
\beq
\frac{1}{2}\sum_{n=-\infty}^{\infty}\(2\omega_n-\omega_{n+m}-\omega_{n-m}\) \nn
\eeq
with $\omega_n=\omega_{-n}$ and assume that for large mode number,
$\omega_n\simeq |n|+\dots$. Naively this sum is zero if $m$ is integer,
since all terms cancels among each other if we allow renumbering of the
terms. However a more careful analysis shows that this is not the case
\beq
\frac{1}{2}\sum_{n=-N}^{N}\(2\omega_n-\omega_{n+m}-\omega_{n-m}\)=
\sum_{n=N-m+1}^{N}(\omega_n-\omega_{n-m}) = m^2 +\O\(1/N\)\, . \nn
\eeq
Thus one should be very careful calculating 1-loop shift having a
frequencies at hand because ambiguities can easily arise.
Consider, for example, the equation for the bosonic frequencies for the general $3\J$ solution \cite{Arutyunov:2003za}
\beqa
P_8^{J_1J_2J_3}(\omega)=\(\omega^2-n^2\)^4-4\(\omega^2-n^2\)^2
\sum_{i\neq j}^3
\frac{\J_i}{w_i}\(w_j \,\omega-m_j \,n\)^2 \nn \\ +\,8 \sum_{i\neq j
\neq k \neq i}^3\frac{\J_i}{w_i}\(w_j \,\omega-m_j \,n\)^2\(w_k
\,\omega-m_k \,n\)^2=0 \,. \nn
\eeqa
This solution can be smoothly deformed to
general $SU(2)$ solution for which $J_3\rightarrow 0$ while preserving all constraints (\ref{const}). In this limit
we find
\beq
  P_8^{J_1J_20}(\omega)=P_4^{su(2)}(\omega)\(\(\omega^2-n^2\)^2-4\(m_3 n-\omega\sqrt{m_3^2+\nu^2}\)^2\) \nn
\eeq
where the quartic polynomial $P_4^{su(2)}(\omega)$ is the one appearing in table 3 and gives us the usual $SU(2)$ modes whereas the
remaining equations yields the frequencies
\beq
\sqrt{(n+m_3)^2+\nu^2}+w_3,\;\;\;\;\;\sqrt{(n-m_3)^2+\nu^2}-w_3 \nn
\eeq
instead of the two $\sqrt{n^2+\nu^2}$ we read from table 3.
From the above explanation this ambiguity converts into an extra contribution
of $m_3^2/\kappa$ to the 1-loop shift.

Moreover, we also found contradictory results in the literature. For the simple $SU(2)$ solution,
in \cite{Frolov:2003tu,Frolov:2004bh,Schafer-Nameki:2006gk} the sum
over fermionic frequencies $\omega^F_n$ is taken over the integers for
even $m$ and over $\mathbb{Z}+1/2$ for odd $m$ while in
\cite{Beisert:2005mq,Hernandez:2006tk} the sum always goes over the integers.
We found that the fermions will indeed be
summed over integer $n$'s.
The same kind of mismatch appears for the $sl(2)$ circular string.
For example, in \cite{Schafer-Nameki:2005tn,Beisert:2005cw,Hernandez:2006tk,Park:2005ji}
the fermionic  frequencies $\omega_n^F$ are summed with $n$ integer whereas we found
 $\omega^F_{n+m/2-k/2}$ and $\omega^F_{-n-m/2-k/2}$ that is, the frequencies have half--integer arguments
 if $m+k$ is odd -- see (\ref{Esl2}).
In view of this discrepancies we also repeated the calculation for the frequencies directly from the expansion
of the string action using a coset representative parameterized as in \cite{Alday:2005jm}. We also found the
same kind of field redefinitions which are trivially related to the $\mathcal{R}$ and $\mathcal{Q}$ matrices
written in section \ref{circular}. For the simple $SU(2)$ solution the field redefinitions always leave the fermions
periodic whereas for the $sl(2)$ string they are periodic (anti--periodic) for $m+k$ even (odd) in agreement with
the calculation presented in this paper.
Shifts changing integers into half-integers are no longer are related by the simple expressions of the form
$m^2/\kappa$ like in the previous example.
However, the sum over fermions can be replaced integral with
exponential precision and therefore this shifts may end up being not so harmful.

\end{document}